\documentclass[12pt]{article} %***
\usepackage[sectionbib]{natbib}
\usepackage{array,epsfig,fancyheadings,rotating}
\RequirePackage{amsthm,amsmath}
\usepackage{hyperref} 
\usepackage{xr}
\usepackage{setspace}
\usepackage[usenames, dvipsnames]{color}
\usepackage{showlabels}
\usepackage{graphicx}
\usepackage{bm}
\usepackage{float}
\usepackage{mathrsfs}
\usepackage[font={small,it}]{caption}
\usepackage{bbold}
\usepackage{multirow}
\usepackage{bbm}
\usepackage{chngpage}
\usepackage{adjustbox}
\usepackage{nicefrac}
\usepackage{amssymb}
\usepackage{bigints}

\newcommand{\Real}{\mathbb{R}}

\usepackage{sectsty, secdot}
\sectionfont{\fontsize{12}{6pt plus.8pt minus .6pt}\selectfont}
\renewcommand{\theequation}{\thesection\arabic{equation}}
\subsectionfont{\fontsize{12}{6pt plus.8pt minus .6pt}\selectfont}
\usepackage{amsmath}
\usepackage{amssymb}
\usepackage{amsfonts}
\usepackage{multirow}
\usepackage{amsthm}

\newtheorem{theorem}{Theorem}

\newtheorem{condition}[theorem]{Condition}

\newtheorem{result}[theorem]{Result}

\theoremstyle{definition}

\begin{document}

%%%%%%%%%%%%%%%%%%%%%%%%%%%%%%%%%%%%%%%%%%%%%%%%%%%%%%%%%%%%%%%%%%%%%%%%%%%%%%%%%%%%%%%%%%%%%%%%%%%%%%%%%%%%%%%%%%%%%%%%%%%%
%%%%%%%%%%%%%%%%%%%%%%%%%%%%%%%%%%%%%%%%%%%%%%%%%%%%%%%%%%%%%%%%%%%%%%%%%%%%%%%%%%%%%%%%%%%%%%%%%%%%%%%%%%%%%%%%%%%%%%%%%%%%

\renewcommand{\baselinestretch}{2}

\markright{ \hbox{\footnotesize\rm 
%{\footnotesize\bf 24} (201?), 000-000
}\hfill\\[-13pt]
\hbox{\footnotesize\rm
%\href{http://dx.doi.org/10.5705/ss.20??.???}{doi:http://dx.doi.org/10.5705/ss.20??.???}
}\hfill }

\markboth{\hfill{\footnotesize\rm Sara Algeri and David A. van Dyk} \hfill}
{\hfill {\footnotesize\rm Testing One Hypothesis Multiple Times} \hfill}

\renewcommand{\thefootnote}{}
$\ $\par

%%%%%%%%%%%%%%%%%%%%%%%%%%%%%%%%%%%%%%%%%%%%%%%%%%%%%%%%%%%%%%%%%%%%%%%%%%%%%%%%%%%%%%%%%%%%%%%%%%%%%%%%%%%%%%%%%%%%%%%%%%%%

\fontsize{12}{14pt plus.8pt minus .6pt}\selectfont \vspace{0.8pc}
\centerline{\large\bf Testing One Hypothesis Multiple Times}
\vspace{.4cm} \centerline{Sara Algeri$^{*,\dag}$ and David A. van Dyk$^{\dag}$} \vspace{.4cm} \centerline{\it
 \textcolor{black}{$^{*}$University of Minnesota} 
and $^{\dag}$  Imperial College London} \vspace{.55cm} \fontsize{9}{11.5pt plus.8pt minus
.6pt}\selectfont

%%%%%%%%%%%%%%%%%%%%%%%%%%%%%%%%%%%%%%%%%%%%%%%%%%%%%%%%%%%%%%%%%%%%%%%%%%%%%%%%%%%%%%%%%%%%%%%%%%%%%%%%%%%%%%%%%%%%%%%%%%%%

\begin{quotation}
\noindent {\it Abstract:}
In applied settings, tests of hypothesis where a nuisance parameter  is only identifiable under the alternative often reduces  into one of  \emph{Testing One Hypothesis Multiple times} (TOHM). Specifically, a fine discretization of the space of the non-identifiable parameter  is specified,  and  the
null hypothesis is tested against  a set of  \emph{sub-alternative hypothesis}, one for each point of the discretization.  The resulting \emph{sub-test statistics} are then combined  to obtain a \emph{global p-value}.
In this paper, we  discuss a  computationally efficient inferential tool to perform TOHM under stringent significance requirements, such as those typically required in the physical sciences, (e.g., p-value $<10^{-7}$).  
The resulting procedure leads to a generalized approach to perform inference under non-standard conditions, including non-nested models comparisons.

\vspace{9pt}
\par
\end{quotation}\par

\def\thefigure{\arabic{figure}}
\def\thetable{\arabic{table}}
\renewcommand{\theequation}{\thesection.\arabic{equation}}
\fontsize{12}{14pt plus.8pt minus .6pt}\selectfont
%%%%%%%%%%%%%%%%%%%%%%%%%%Section 1
\section{Introduction}
\label{intro}

A fundamental statistical challenge in scientific discoveries is the so called ``bump-hunting" problem \citep{bumpphy}, where researchers aim to distinguish  peaks due to a signal of interest (the new discovery) from peaks due to   random fluctuations of the background. In the framework of   hypothesis testing, the null model specified by $H_0$ is typically the background-only model, and a signal bump is   added in the alternative model specified by $H_1$.
Consider for example a dark matter search where we aim to distinguish events associated with a power-law (Pareto type~I) distributed   background from the signal of a dark matter source modeled as a narrow Gaussian bump with unknown location \textcolor{black}{over the search area $\Theta\equiv[\mathcal{L},\mathcal{U}]\subset \Real$}.  We can specify the model of interest using a mixture model
\begin{equation}
\label{ex1}
(1-\eta)\frac{1}{k_{\phi}y^{\phi+1}}+ \frac{\eta }{k_{\theta}}\exp\biggl\{-\frac{(y-\theta)^2}{0.02\theta^2}\biggl\}\qquad \text{for $y\geq1$,}
\end{equation}
where $k_{\phi}$ and $k_{\theta}$  are normalizing constants, $y\geq1$, $\phi>0$, and $\theta\geq1$.  \textcolor{black}{Notice that   the parameter $\theta$  characterizes both  the location of the signal over the search region and    its standard deviation. Specifically,  the bump becomes   wider the further   its position is in the tail of the background distribution. The model in \eqref{ex1} is a toy example which simplifies the models involved in the context of  searches for $\gamma$-ray emissions ina  cluster of galaxies \citep{refB3}; where for example the width of the signal may be a more complex function of its location. Despite its simplicity, the model in \eqref{ex1} introduces the key statistical issues arising in  the context of dark matter searches, as described below.}

In order to assess the evidence in favor of the  signal, we test 
\begin{equation}
\label{testex1}
H_0: \eta=0 \quad \mbox{versus} \quad H_1:\eta>0.
\end{equation}
where $\eta$ is the proportion of events due to the dark matter emission, and typically $0\leq \eta \leq 1$.
Despite its straightforward formulation, testing  \eqref{testex1} is non-trivial. Difficulties arise because   $\theta$   is not defined under $H_0$. Consequently, classical asymptotic properties of, e.g.,  Maximum Lixelihood Estimates (MLE) and the Likelihood Ratio Test (LRT), fail. Analogously, complications may arise  when using  resampling techniques, such as bootstrapping \citep{boostrap}, to derive the null distribution of the test statistic,  in the presence of stringent significance requirements. \textcolor{black}{For searches in high energy physics for instance,  the significance level necessary to claim a discovery can be in the order of   $10^{-7}$ \citep[see][Table 1]{lyons2015}. 
Hence, a large (e.g., $O(10^8)$) simulation may be infeasible when dealing with complex models. This is a key motivation for   a computationally efficient inferential solution}. %: a limit for the MLE is not guaranteed to exist and  Wilk's theorem \citep{wilks} cannot be applied to derive the asympotic distribution of the LRT. 

 \textcolor{black}{To address these difficulties,} in this paper, we consider the bump-hunting problem as a  special case of what is known in statistical literature as \emph{``testing statistical  hypotheses when a nuisance parameter is present only under the alternative''}. % or equivalently, \emph{``testing with unidentifiable parameters under $H_0$''}. 
In addition to   bump-hunting, classical examples may  include regression models where structural changes, such as break-points and  threshold-effects, occur \citep{andrews93, hansen92b, hansen99,davies02}. 

The general problem has long been studied, starting at least from the seminal work of \citet{hotelling39} and \citet{davies77,davies87}, and further investigated in the econometrics literature  by several authors including \citet{andrews94} and \citet{hansen91,hansen92,hansen96}.  
In their practical implementation, these methods reduce the problem of testing with unidentifiable parameters under $H_0$ into one  of \emph{Testing One Hypothesis Multiple times} (TOHM),  where a single null hypothesis $H_0$ is tested against  different \emph{sub-alternative hypotheses} of the form $H_{1}(\theta)$, one for each fixed $\theta$ in $\Theta$, and a corresponding ensemble of \emph{sub-test statistics} indexed by $\theta$, namely $W(\theta)$, is specified. 
The goal is to provide a global p-value as the standard of evidence for comparing $H_0$ and the global alternative hypothesis $H_1$, of which each   $H_{1}(\theta)$   is a  special case. 
Unfortunately, existing methods often require case-by-case mathematical computations \citep[e.g.,][]{davies77}, estimating the covariance structure  \citep[e.g.,][]{hansen91}, choosing weighting functions  \citep[e.g.,][]{andrews94}, or full  simulations of the empirical process  \citep[e.g.,][]{hansen92, hansen96}. 

 In this paper we discuss a computationally efficient method to perform TOHM  which overcomes these limitations. 
Specifically, as in \citet{davies77,davies87} \textcolor{black}{we consider a stochastic process, $\{W(\theta)\}$,  indexed by $\theta\in\Theta\equiv[\mathcal{L},\mathcal{U}]$, }  and  with covariance function $\rho(\theta,\theta^\dag)$. We consider the global p-value 
\begin{equation}
\label{pval}
P\biggl(\sup_{\theta \in \Theta}\{W(\theta)\}>c\biggl),
\end{equation}
where $c$ is the observed value of the \emph{global test statistic}, $\sup_{\theta \in \Theta}\{W(\theta)\}$. 
The central difficulty of this approach is  to derive or approximate \eqref{pval}. One possible way forward is to consider the Extreme Value Theory (EVT) argument developed by  \citet[p. 272]{cramer}, where a bound for   \eqref{pval} is obtained  considering the upcrossings of $c$ by $\{W(\theta)\}$ (see Figure~\ref{exceedances}). 
Specifically,   $\{W(\theta)\}$  has an \emph{upcrossing} of a threshold $c\in \Real$ at $\theta_0 \in \Theta$ if, for some $\epsilon>0$, $W(\theta)\leq c$ in the interval $(\theta_0-\epsilon,\theta_0)$ and $W(\theta)\geq c$ in the interval $[\theta_0,\theta_0+\epsilon)$ \citep{adler2000}. 
Let $N_c$ be the number of upcrossings of $c$ by $\{W(\theta)\}$. Using Markov's inequality, \citet[p. 272]{cramer} show that \eqref{pval} can be bounded as in \eqref{general_bound},
\begin{equation}
\label{general_bound}
P\biggl(\sup_{\theta \in \Theta} \{W(\theta)\}>c\biggl)\leq P(W(\mathcal L)>c)+E[N_c]
\end{equation}
 where $P(W(\mathcal L)>c)$ is typically known. %, whereas, $E[N_c]$  bounds $P(N_c\geq 1)$. 
\citet{davies77, davies87} consider the cases where $\{W(\theta)\}$ is a Gaussian or a $\chi^2$-process, estimate $E[N_c]$  via total variation, and show  that \eqref{general_bound} becomes sharp,  as $c\rightarrow \infty$ (under long-range independence, i.e., if $\rho(\theta,\theta^\dag)\rightarrow 0$ as $|\theta-\theta^\dag|\rightarrow \infty$). 
Unfortunately, \citet{hansen91} points out that  situations exist where the total variation  diverges. 

An alternative solution   can  overcome this problem  and  has had significant impact in  physics \citep{gv10}. Consider a set of observations $y_1,\dots,y_n$, and let $T_n(\theta)$ the LRT statistics used to test \eqref{testex1} and evaluated on  $y_1,\dots,y_n$ when $\theta$ is fixed.
\textcolor{black}{We denote  the LRT-process indexed by different values of $\theta$ with $\{T_n(\theta)\}$}. Under 
 $H_0$ and suitable uniformity conditions \citep{hansen91},   $\{T_n(\theta)\}\xrightarrow[n\rightarrow \infty]{d}\{W_\chi(\theta)\}$,  where $W_\chi(\theta)$ is a $\chi^2$-process with components  $W_\chi(\theta)\sim\chi^2_s$, for each $\theta \in [\mathcal{L,\mathcal{U}]}$ fixed. Let $E[N^\chi_c]$ be the expected number of upcrossings of $c$ by $\{W_\chi(\theta)\}$ over $\Theta$.  One possible way to compute  \eqref{general_bound} is to estimate $E[N^\chi_c]$ via Monte Carlo simulations.  However,  when dealing with stringent significance requirements,  the corresponding significance threshold $c$ is typically very large. Hence,  upcrossings of $c$ are expected to occur infrequently when simulating under $H_0$, and thus a massive simulation is  required to estimate $E[N^\chi_c]$ directly. 
 \citet{gv10} exploit  the $\chi^2$ distribution of  $\{W_\chi(\theta)\}$, and rewrite $E[N^\chi_{c}]$ as a function of $E[N^\chi_{c_0}]$, see \eqref{gv_bound}, for  some $c_0<<c$,
\begin{equation}
\label{gv_bound}
P\biggl(\sup_{\theta \in \Theta} \{W_\chi(\theta)\}>c\biggl) \leq P(W_\chi(\mathcal{L})>c)+\biggl(\frac{c}{c_0}\biggl)^{\frac{s-1}{2}}e^{-\frac{c-c_0}{2}}E[N^\chi_{c_0}].
\end{equation}
where $E[N^\chi_{c}]=\biggl(\frac{c}{c_0}\biggl)^{\frac{s-1}{2}}e^{-\frac{c-c_0}{2}}E[N^\chi_{c_0}]$.
 This allows a drastic reduction in  the computational effort needed to compute $E[N^\chi_c]$. Specifically  upcrossings of $c_0<<c$ are expected to occur often, and thus $E[N^\chi_{c_0}]$ can be estimated accurately with a small  Monte Carlo simulation. 

\citet{gv10} do not formally justify \eqref{gv_bound}. In Section~\ref{sec3}, we  derive  \eqref{gv_bound}, we generalized it to any process $\{W(\theta)\}$, and we clarify the conditions under which \eqref{gv_bound} and its generalization hold. Efficient choices of $c_0$ are discussed in Section  \ref{practice} and a simple graphical tool is proposed  to validate the adequacy of the number of sub-tests conducted. 

The resulting procedure leads to a generalized approach to perform inference under non-standard regularity conditions including, as discussed in Section~\ref{practice},  comparisons of non-nested models. This can be done by specifying a  comprehensive model  that includes the two (non-nested) models under comparison as special cases. Two tests of hypothesis where a nuisance parameter is present only under the alternative are then performed to select among the two models \citep{algeri16}. 

In principle, the problem of testing in presence of a nuisance parameter which is present only under the alternative can be formulated as a multiple hypothesis testing (MHT) problem, where several tests are conducted over a grid of possible values of $\theta$, and corrected using Bonferroni's correction \citep{bonferroni35, bonferroni36} or similar methods  to control for the probability of type I error.  \textcolor{black}{Although the Bonferroni correction is easy to implement, it is often dismissed by practitioners both because of its stringent  control of the overall false detection rate and its artificial dependence on the number of tests conducted. In Section~\ref{analysis} we compare TOHM and Bonferroni's correction via a suite of numerical studies and data applications; we also discuss how the tools introduced in this manuscript can be used to identify situations where, by virtue of its relationship with TOHM, Bonferroni can be used without worry about obtaining an overly conservative result. }

The remainder of the paper is organized as follows. 
In Section~\ref{sec3},  we define the framework for TOHM, and we derive a computable upper bound for \eqref{pval} by generalizing  \eqref{gv_bound}. In Section~\ref{practice},  we illustrate how TOHM can be used to distinguish among non-nested models, \textcolor{black}{we validate our results with simulation studies} and we discuss graphical tools \textcolor{black}{to select the  necessary quantities involved in the computation of the bound proposed   in Section \ref{sec3}.  In Section \ref{analysis} we investigate the relationship between TOHM and the classical Bonferroni correction,  and  we apply both methods on several realistic data sets. A summary and a discussion of our findings appear in Section~\ref{discussion}. 
Additional figures, data and proofs   are collected in the Supplementary Material.}

%%%%%%%%%%%%%%%%%%%%%%%%%%%Section 2

%%%%%%%%%%%%%%%%%%%%%%%%%%

\section{TOHM via EVT}
\label{sec3}
\subsection{Definition and formalization}
\label{GV}
In this section, we generalize the testing procedure of \citet{gv10} beyond the LRT   and the $\chi^2$ case and formalize it in statistical terms.  This allows us to establish a general theoretical framework to efficiently bound/approximate the global p-value in \eqref{pval}. 

Recall  that  $\{W(\theta)\}$ is a generic stochastic process indexed by $\theta \in \Theta \equiv [\mathcal L; \mathcal U]$ with covariance function $\rho(\theta,\theta^\dag)$. Following \citet{davies87} we stipulate 
\begin{condition}
\label{cond31}
$\{W(\theta)\}$ has continuous sample paths;
$\{W(\theta)\}$ has continuous first derivative, except possibly for a finite number of jumps;
and  its components $W(\theta)$ are identically distributed for all $\theta \in \Theta$.
\end{condition}

To exploit \eqref{general_bound}, we aim to conveniently estimate $E[N_c]$  and bound or approximate \eqref{pval}.  Results \ref{theo1} and \ref{coroll1} allow this.
\begin{result}
\label{theo1}
Let  $c\in \Real$ be an arbitrary threshold, $a(c)$ be a function which depends on $c$ but not on $\theta$, and $b(\Theta)$  be a function which does not depend on $c$, and to be calculated over $\Theta$. 
Under Condition \ref{cond31}, if  $E[N_c]$ can be decomposed as 
\begin{equation}
\label{decompose}
E[N_c]=a(c)b(\Theta)
\end{equation}
then,
 \begin{equation}
\label{expect}
E[N_c]=\frac{a(c)}{a(c_0)}E[N_{c_0}]\qquad \forall c_0\leq c, c_0 \in \Real.
\end{equation}
\end{result}

\textcolor{black}{The function $b(\Theta)$ typically involves  integration over the interval $\Theta$, and should not be confused with a function  of  $\theta$.}
Deriving a closed-form expression of $b(\Theta)$ in \eqref{decompose} may be challenging, and  may require knowledge of $\rho(\theta,\theta^\dag)$. Conversely, the form of $a(c)$  typically  depends on the marginal distribution of the components $W(\theta)$ of $\{W(\theta)\}$, hence the requirement of identical distribution in Condition \ref{cond31}. The continuity assumptions on $\{W(\theta)\}$ and its first derivative prevent $E[N_c]$ from diverging.

Equation \eqref{expect} offers a simple way to compute $E[N_c]$, provided that, as discussed below, $E[N_{c_0}]$ can be estimated accurately. Result \ref{coroll1} follows from \eqref{general_bound}, \eqref{decompose}, and  \eqref{expect}. 
\begin{result}
\label{coroll1}
Under Condition \ref{cond31},  \textcolor{black}{ if \eqref{decompose} holds}, \eqref{pval} can be bounded by
\begin{equation}
\label{bound2}
P\biggl(\sup_{\theta \in \Theta} \{W(\theta)\}>c\biggl)  \leq P(W(\mathcal L)>c)+ \frac{a(c)}{a(c_0)}E[N_{c_0}]
\end{equation}
for all $ c_0\leq c, c_0 \in \Real$. If additionally,  $\rho(\theta,\theta^\dag)\rightarrow 0$ as $|\theta-\theta^\dag|\rightarrow \infty$,  \textcolor{black}{the difference between the left and the right hand side of \eqref{bound2} approaches zero as $c\rightarrow \infty$. }
\end{result}
%In Section~\ref{derivations} of the Supplementary Materials we derive explicit forms of \eqref{bound2} for the   normal, $\chi_s^2$, $\bar{\chi}^2_{01}$ \citep{lin,takemura97}, $F$ and  $t$ cases. The rates of convergence are discussed in Section~\ref{appA}. 

\textcolor{black}{
\subsection{TOHM bounds for Gaussian-related processes}
\label{bounds}
The bound  in \eqref{gv_bound} and  the analogous bounds for   Gaussian and  related processes such as   $F$ and  $t$-processes, can  be derived using results of random fields theory as discussed in \citet{algeri18}. 
In this setting, it can be shown that, under mild smoothness conditions (see \citet[p. 547]{taylor2003}), $E[N_c]$ enjoys the decomposition in \eqref{decompose}, where $a(c)$  only depends on the distribution of the marginals of $\{W(\theta)\}$, whereas $b(\Theta)$ corresponds to the so-called Lipschitz-Killing curvature of first order \citep[e.g.,][]{adlerbook} and  is typically difficult to compute. 
Here, we report  explicit forms of the right hand side of \eqref{bound2} for Gaussian, $F$ and  $t$  processes which can be obtained on the basis of these results  \citep[see][for more details]{taylor2008,adlerbook,algeri18}.}

\textcolor{black}{
\paragraph{Gaussian process.} Let $\{Z(\theta)\}$ be a mean zero and variance one Gaussian process, such that $Z(\mathfrak{\theta})\sim N(0,1)$ for all $\theta\in \Theta$, and let $N^Z_{c}$ be the process of upcrossings of $c_0$ by $\{Z(\theta)\}$ over $\Theta\equiv[\mathcal{L},\mathcal{U}]$.  The TOHM bound in equation \eqref{bound2} takes the form
\begin{equation}
\label{z_bound}
P\biggl(\sup_{\theta \in \Theta} \{Z(\theta)\}\geq c\biggl) \leq \Phi(-c)+e^{-\frac{c^2-c_0^2}{2}}E[N^Z_{c_0}].
\end{equation}
where $\Phi(-c)$ is the cumulative density function of a standard normal random variable evaluated at $-c$ and the   ratio  $\frac{a(c)}{a(c_0)}$ is givan by $e^{-\frac{c^2-c_0^2}{2}}$.
For the stationary case, the same result can be obtained by expressing $E[N^Z_{c}]$ via Rice's formula \citep[]{rice} i.e.,
\[
E[N^Z_{c}]=\frac{|\mathcal{L}-\mathcal{U}|}{2\pi}\sqrt{\rho''(\theta,\theta)}e^{-\frac{c^2}{2}}
\]
where $\rho''(\theta,\theta)=\frac{\partial \theta}{\partial \theta \partial \theta^\dag}\rho(\theta,\theta^\dag)\bigl |_{\theta^\dag=\theta}$ is the second spectral moment of $\{Z(\theta)\}$ and is assumed to be finite, and $|\mathcal{L}-\mathcal{U}|$ is the length of $\Theta$.
As discussed in \citet{davies87}, for   a two-sided test,  the excursion probability of interest is $P(\sup_{\theta \in \Theta} |\{Z(\theta)\}|\geq c)$; the bound  of which is twice the right hand side of  \eqref{z_bound}.}

\textcolor{black}{
The rate of convergence of the difference between the right and left hand side of \eqref{gv_bound} and \eqref{z_bound}    are discussed in  Section S.1 of the Supplementary Material.
We  further study   the sharpness of the bounds in \eqref{gv_bound}  and \eqref{z_bound}, as $c\rightarrow\infty$  in Section~\ref{practice} via a suite of simulation studies. }

\textcolor{black}{
\paragraph{$F$-process.} Consider an $F$-process $\{F(\theta)\}$ with $s$ and $v$  degrees of freedom  such that $F(\theta)\sim F_{s,v}$ for all $\theta\in \Theta$. Let $E[N^F_{c_0}]$ be the expected number of upcrossings of $c_0$ by $\{F(\theta)\}$, then the TOHM bound in equation \eqref{bound2} takes the form
\begin{equation}
\label{F_bound}
P\biggl(\sup_{\theta \in \Theta} \{F(\theta)\}\geq c\biggl)  \leq P(F(\mathcal L)\geq c)+ \biggl(\frac{c}{c_0}\biggl)^{\frac{s-1}{2}}\biggl(\frac{v+s\cdot c}{v+s\cdot c_0}\biggl)^{-\frac{s+v-2}{2}}E[N^F_{c_0}] 
\end{equation}
for all $c_0\leq c, c_0 \in \Real$, and with $a(c)=c^\frac{s-1}{2}(v+s\cdot c)^{-\frac{s+v-2}{2}}$.}

\textcolor{black}{
\paragraph{$t$-process.} Consider a $t$-process $\{V(\theta)\}$ with $s$ degrees of freedom   such that $V(\theta)\sim t_{s}$. Let $E[N^V_{c_0}]$ be the expected number of upcrossings of $c_0$ by $\{V(\theta)\}$, then the TOHM bound in equation  \eqref{bound2} takes the form
\begin{equation}
\label{t_bound}
P\biggl(\sup_{\theta \in \Theta} \{V(\theta)\}\geq c\biggl)  \leq P(V(\mathcal L)\geq c)+ \biggl(\frac{1+c^2}{1+ c^2_0}\biggl)^{-\frac{s-1}{2}}E[N^V_{c_0}] 
\end{equation}
for all $c_0\leq c, c_0 \in \Real$, and with $a(c)=(1+ c^2)^{-\frac{s-1}{2}}$.}

\textcolor{black}{\subsection{Testing one hypothesis multiple times in practice}
\label{Ncrdef}}

In practice,  we evaluate $\{W(\theta)\}$ on a fine grid of points, namely $\Theta_R=\{\theta_1, \dots\theta_R\}  \subseteq \Theta$, with $R$ being the typically large number of grid points. 
Let $\{W(\theta_r)\}$ be the random sequence which coincides with $\{W(\theta)\}$ at each  $\theta_r\in\Theta_R$ and $\{w(\theta_r)\}$ be its observed value. We approximate  $\sup_{\theta \in \Theta}\{W(\theta)\}$  with its discrete counterpart $\max_{\theta_r \in \Theta_R}\{W(\theta_r)\}$, the observed value of which is given by
\begin{equation}
\label{cR}
c_R=\max_{\theta_r \in \Theta_R}\{w(\theta_r)\}.% \qquad \text{such that} \qquad c=\lim_{R\rightarrow\infty } c_R<\infty.
\end{equation}
Let  the process of  upcrossings  of $c_R$ by $\{W(\theta_r)\}$, namely $\tilde{N}_{c_R}$, be events of the type  $\{W(\theta_{r-1})\leq c_R, W(\theta_r)>c_R\}$. 
We assume that $\Theta_R$ is sufficiently dense, so that the right hand side of \eqref{bound2} can be approximated by \eqref{real_bound}, as $R\rightarrow \infty$,
\begin{equation}
\label{real_bound}
P(W(\mathcal L)>c_R)+ \frac{a(c_R)}{a(c_0)}E[\tilde{N}_{c_0}]\qquad \forall c_0\leq c_R, c_0 \in \Real
\end{equation} 
where $E[\tilde{N}_{c_0}]$ can be replaced by its Monte Carlo estimate, namely $\widehat{E[\tilde{N}_{c_0}]}$. 

\textcolor{black}{Notice that the null hypothesis, $H_0$, is tested versus an ensable of alternative hypotheses $H_{1r}$, one for each value of $\theta_r$ fixed. The observed \emph{sub-test statistics} $\{w(\theta_1),\dots,w(\theta_R)\}$, realizations of $\{ W(\theta)\}$,  are  combined into the global test statistic $c_R$ and  an approximated bound for the  global p-value is computed via \eqref{real_bound}. Thus, the problem of testing \eqref{testex1} is  reduced to testing $H_0$ versus the $R$  \emph{sub-alternative hypotheses}  $H_{1r}$,  i.e., \emph{Testing One Hypothesis Multiple Times}. }

\citet[p. 63 and 195]{cramer}  discuss adequate choices of $\Theta_R$ for which $c$, $N_{c}$ and  $\sup_{\theta \in \Theta}\{ W(\theta)\}$ are well approximated by $c_R$, $\tilde{N}_{c_R}$ and   $\max_{\theta_r \in \Theta_R}\{ W(\theta_r)\}$, respectively. However, since in practice $\Theta_R$ may be determined by the experiment, in Section~\ref{practice} we discuss  graphical tools to assess whether these approximations hold. 

%In general, the threshold $c_0$ should be chosen small enough to catch a reasonably high number of upcrossings, but yet high enough to maintain a suitable distance between upcrossings vis-a-vis the resolution of $\Theta_R$.
\vspace{1cm}

\begin{figure}[!h]
\centering
\begin{adjustwidth}{0cm}{0cm}
\begin{tabular*}{\textwidth}{@{\extracolsep{\fill}}@{}c@{}c@{}c@{}}
      \includegraphics[width=45mm]{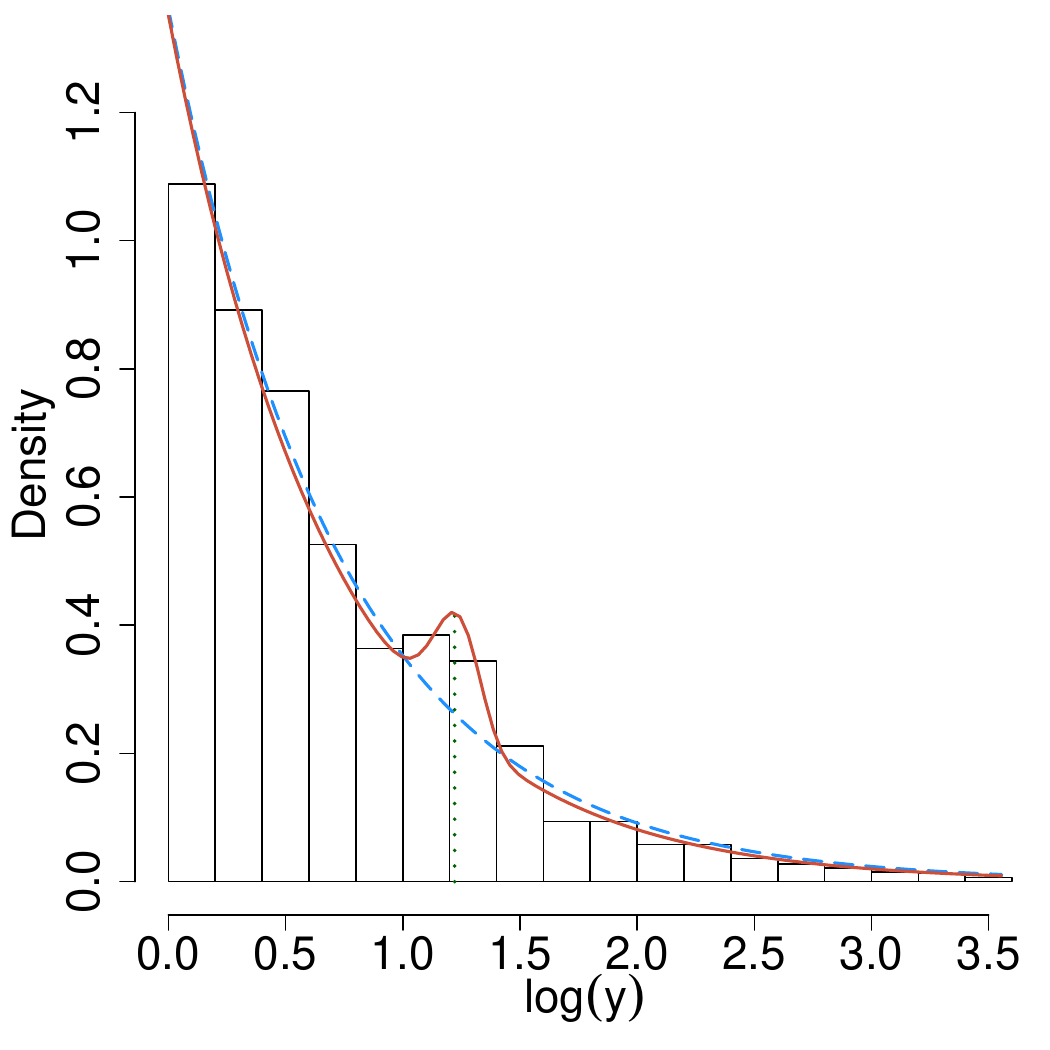} & \includegraphics[width=45mm]{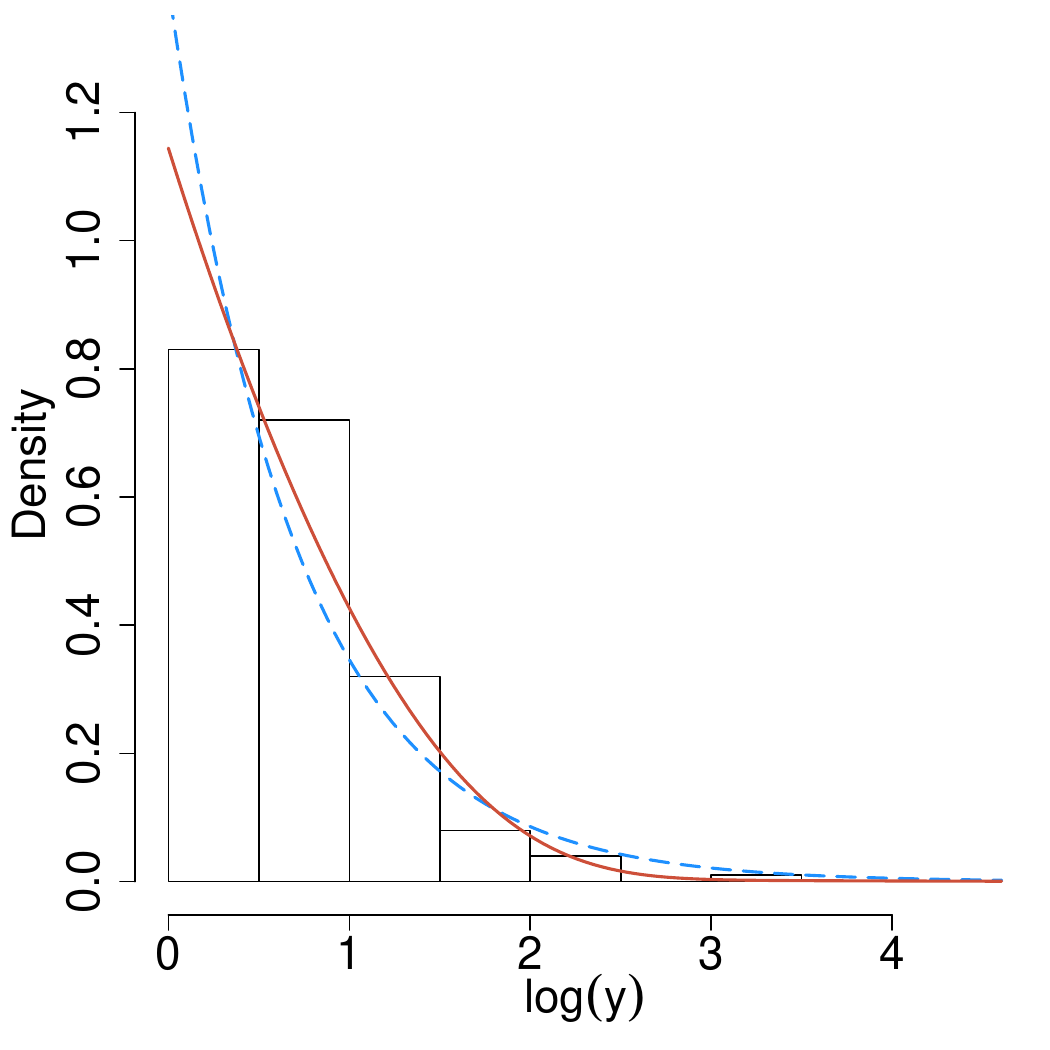}& \includegraphics[width=45mm]{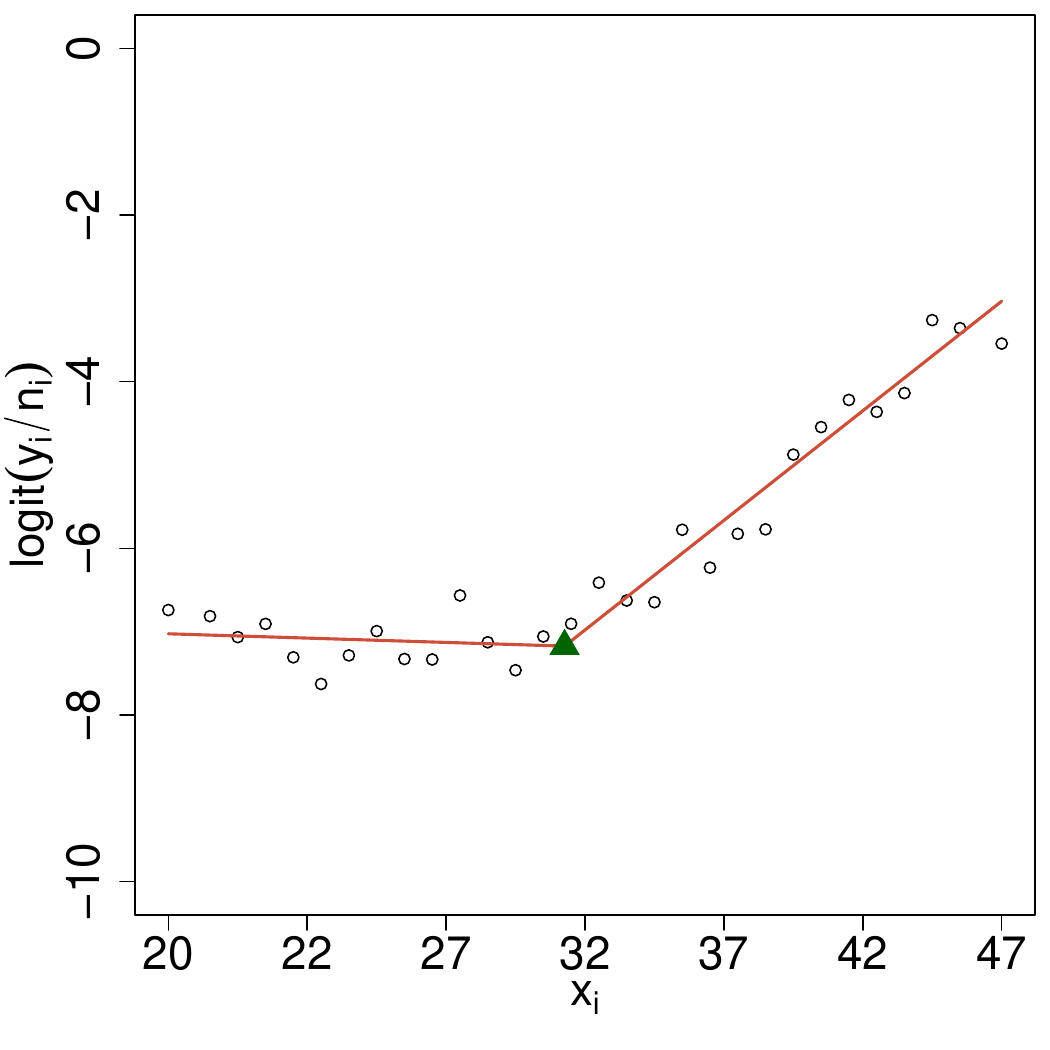} \\
\end{tabular*}
\end{adjustwidth}
\caption{Data and fitted models. Left panel: histogram of the Fermi-LAT realistic data simulation for Example~1 (on $log$-scale), null model (blue dashed curve) fitted  under the assuption of background only counts ($\hat{\phi}=1.350$), and fitted alternative model (red solid curve) with $\hat{\eta}=0.045$, $\hat{\phi}=1.406$. The green dotted vertical line indicates the location of the observed Gaussian bump, i.e., $\hat{\theta}=3.404$. Central panel: histogram of the Fermi-LAT realistic data simulation for Example~2 ($log$-scale), \textcolor{black}{the null model when testing \eqref{testex1} is fitted as a power-law distributed cosmic source with $\hat{\phi}=1.395$  (blue dashed curve). The null model when testing \eqref{flipping} is   the dark matter model in \eqref{DMmodel} with $\hat{\theta}=27.89$ obtained via MLE  (red solid curve) }.  Right panel: Down syndrome data and fitted regression model (red piecewise-linear solid lines), with break-point (green triangle) at $\hat{\theta}=31.266$.}
\label{real_plots}
\end{figure}  
\section{Practical matters}
\label{practice}
\subsection{Case studies: description}
\label{examples}
Here we illustrate the implementation of TOHM in the context of three case studies, i.e., the ``bump hunting" problem introduced in Section~\ref{intro}, a  non-nested models comparison, and a logistic model with a break point. Hereafter,  we  refer to these as Examples~1, 2 and 3, respectively. 
Data for Examples~1 and 2 were generated using simulations of the Fermi Large Area Telescope (LAT) 
obtained with the \emph{gtobssim} package\footnote{\url{http://fermi.gsfc.nasa.gov/ssc/data/analysis/software}} and include  representations of detector  effects and systematic errors. The Fermi-LAT is a $\gamma$-ray telescope on the orbiting Fermi satellite \citep{atwood}.

In Example~1, our data analysis aims to properly distinguish between $\gamma$-ray signals induced by dark matter annihilations and those induced by the astrophysical background. As in \eqref{ex1}, 
 dark matter events are modeled as a  Gaussian bump with mean energy $\theta$ and standard deviation varying with $\theta$. 
\textcolor{black}{The astrophysical background is power-law (Pareto type~I) distributed with index $\phi$. In our simulation, we set $\theta=3.5$GeV \textcolor{black}{(where GeV denotes Giga electron-volt)}, $\phi=1.4$, $\eta=0.02$, and we consider the energy band $y \in [1;35]$.} This setup resulted in 64 dark matter events and 2274 background events. For more physics details, see \citet{algeri2}.

In Example~2, the non-nested models to be compared are  a dark matter emission   with probability density given by
\begin{equation}
\label{DMmodel}
g(y, \theta) \propto y^{-1.5}\exp\biggl\{-7.8\frac{y}{\theta}\biggl\},
\end{equation}
with  $y \geq 1$, $\phi>0$ and $\theta\geq 1$ \citep[see][]{bergstrom}  and a  power-law  distributed cosmic source with density $f(y,\phi)\propto \frac{1}{k_{\phi}y^{\phi+1}} $. In our simulation we set the putative dark matter emission to occur at $\theta=35$GeV, and the power-law index to $\phi=1.4$. In this way, we obtained  200 dark matter events over the energy band  $y \in [1;100]$.

Since the models $f(y,\phi)$ and $g(y, \theta)$ are non-nested,  the classical asymptotic properties of the MLE  and  LRT fail.  However, as shown in \citet{algeri16}, the framework of Section~\ref{sec3} can be extended to compare non-nested models by reformulating this comparison as a test   in which a nuisance parameter is identified only under $H_1$. Specifically, following   \citet{cox62} and \citet{atkinson}, we specify a  comprehensive model  that embes two non-nested models, i.e.,
\begin{equation}
\label{comprehensive}
(1-\eta)f(y,{\phi})+ \eta g( y, \theta) \qquad 0\leq\eta\leq1.
\end{equation}
This  reduces the problem to a nested models comparison and  we test  \eqref{testex1}.
However, in contrast to the bump-hunting example in \eqref{ex1}, here $\eta$ has no physical interpretation. Rather, as in \citet{quandt}, $\eta$ is an auxiliary parameter which allows us to exploit the normality of its MLE to apply well-know asymptotic results. 
 In addition to \eqref{testex1}, the hypotheses 
\begin{equation}
\label{flipping}
H_0: \eta=1 \quad \mbox{versus} \quad H_1:\eta<1
\end{equation}
should also be tested in order to exclude intermediate situations \citep[e.g.,][]{cox62,cox13}. I.e., we want to avoid treating \eqref{comprehensive}  as a mixture and focus on comparing the two models. Testing both \eqref{testex1} and \eqref{flipping}  is particularly suited to particle physics searches where researchers typically assign different degrees of belief to the models being tested.  Specifically, as described in \citet{dvd}, the most stringent significance requirements \citep[e.g.,][Table 1]{lyons2015} are typically used only in the \emph{detection} stage, i.e., when testing \eqref{testex1} to assess the presence of a new signal. Conversely, in the \emph{exclusion} stage, i.e., when testing \eqref{flipping} to exclude the hypothesis of a signal being present, a significance level of 0.05 is typically sufficient.
The Fermi-LAT datasets for Examples~1 and 2 are plotted in the first two panels of Figure~\ref{real_plots}. Both simulations  are  downloadable among the Supplementary Materials.

Finally, in Example~3 we consider the \emph{Down Syndrome dataset} available in the R package \verb segmented    \citep{segmented}. The dataset 
records whether babies born to 354,880 women are affected by Down Syndrome. We use \eqref{ex3} to model the probability, $\pi_i$, that a woman of age $x_i$ has a baby with down syndrome, where $x_i\in[17;47]$,  and we let $\theta\in[20;44]$. The logit of the ratio between the number of down syndrome cases and number of births by age group is plotted in the right panel of Figure~\ref{real_plots}. 
\begin{equation}
\label{ex3}
\log\biggl(\frac{\pi_i}{1-\pi_i}\biggl)=\phi_1+\phi_2 x_i+\xi(x_i-\theta)\mathbbm{1}_{\{ x_i\geq\theta \}} \quad \forall i=1,\dots,n,
\end{equation} 
where  $\theta \in \Real$ is the location of the unknown break-point. In this case, we test $H_0:\xi=0$ versus $H_1:\xi \neq 0$.

In Example~1 and 2 we use the LRT, $T_n(\theta)$,  as the sub-test statistic. Since both tests are of the form in \eqref{testex1}, the test is  on the boundary of the parameter space and  \textcolor{black}{for each $\theta$ fixed the asymptotic distribution under $H_0$ is a mixture of $\chi_1^2$ and zero \citep{chernoff, self}, also known as $\bar{\chi}$-distribution and which we dentote with $\bar{\chi}^2_{01}$. It can be shown \citep{algeri18} that in this setting the bound in \eqref{bound2} has the same form as in the $\chi^2_1$ case, i.e., it is given by \eqref{gv_bound} with $s=1$.} In Example~3, we use the signed-root of the LRT $Q_n(\theta)=\text{sign}(\hat{\eta}_\theta-\eta_0)\sqrt{T_n(\theta)}$, hence the sub-tests statistics are  asymptotically  normally distributed under $H_0$ \citep[e.g.,][]{davies77}.

\subsection{The choices of $c_0$ and $R$}
\label{choosingR}

\begin{figure}
\begin{tabular*}{\textwidth}{@{\extracolsep{\fill}}@{}c@{}c@{}}
      \includegraphics[width=60mm]{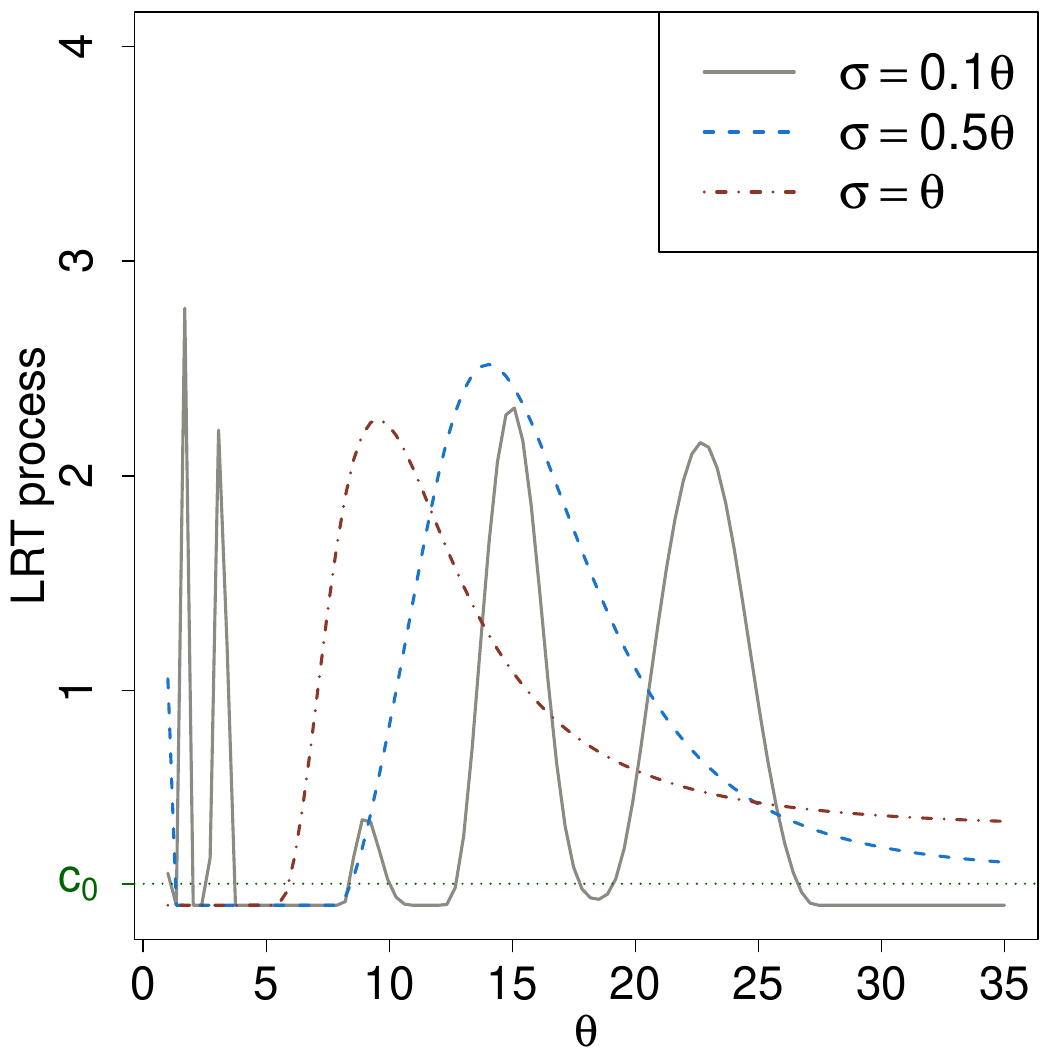} &  \includegraphics[width=60mm]{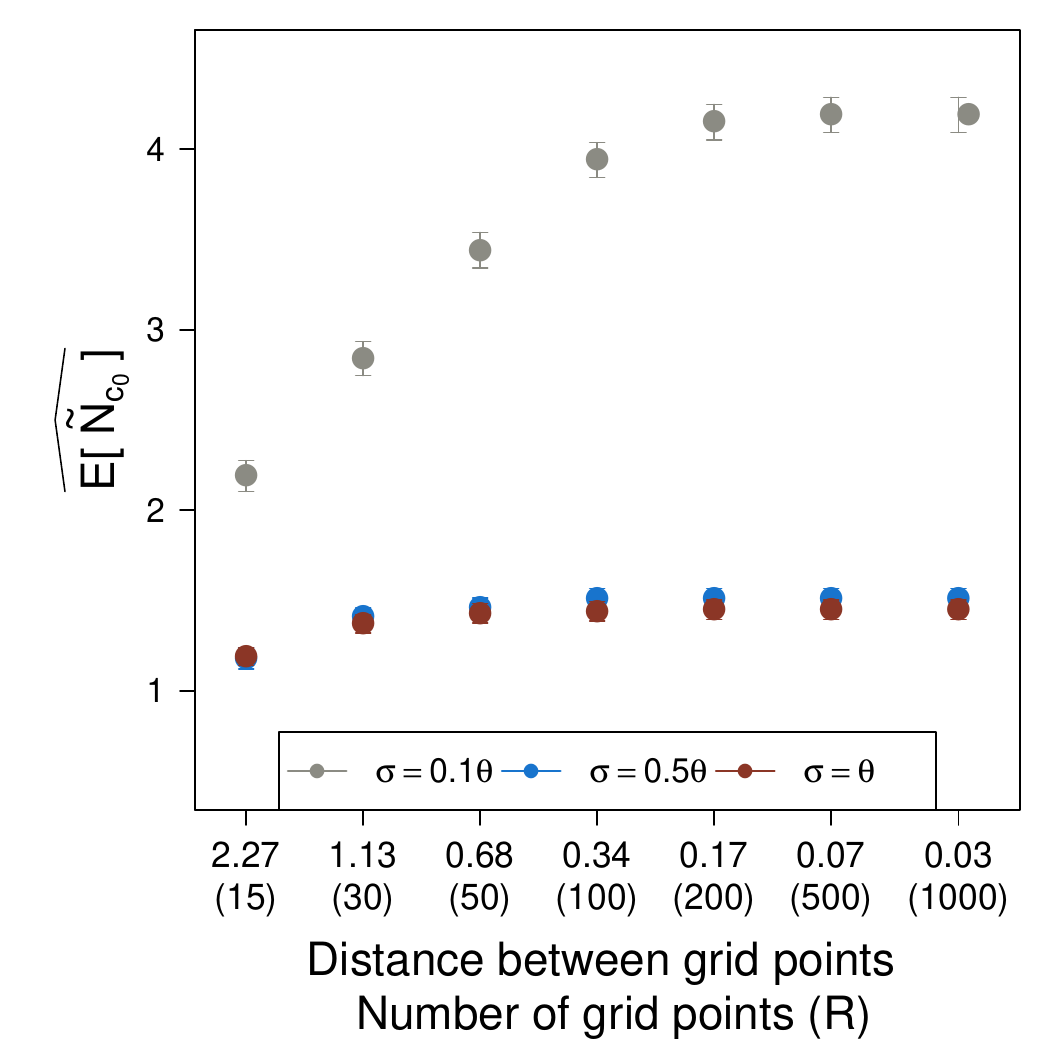} \\
\end{tabular*}
\caption{Left panel:  simulated sample paths  of the LRT process, $\{T_n(\theta)\}$,  under $H_0$ in Example~1. Both plots consider different  widths of the Gaussian bump. Right panel: upcrossings plot showing Monte Carlo estimates of $E[\tilde{N}_{c_0}]$ and standard errors (whiskers), under $H_0$, for Example~1, and evaluated over grids of $R=15,30,50,100,200,500$ points, and for three choices of the Gaussian width, namely $\sigma =0.1\theta, \sigma =0.5\theta$ and $\sigma =\theta$. }
\label{upc_Gauss}
\end{figure}

One way to select an appropriate thresholds $c_0$ is to perform a sensitivity analysis based on few Monte Carlo simulations of  the traces of the underlying processes under $H_0$.
\textcolor{black}{As discussed in Section~\ref{sec3}, under suitable regularity conditions and when $H_0$ is true, the LRT and signed-root LRT processes  $\{T_n(\theta)\}$ and $\{Q_n(\theta)\}$ converge uniformly to $\{W_\chi(\theta)\}$ and $\{Z(\theta)\}$, respectively, as $n\rightarrow+\infty$. More generally, given a test statistics $W_n(\theta)$ to be evaluated on the data $y_1,\dots,y_n$ for each $\theta$ fixed, we write $\{W_n(\theta)\}\xrightarrow{d}\{W(\theta)\}$. Consequently,  for each sample generated under $H_0$, we  compute  $\{W_n(\theta)\}$  over a fine grid of values of $\theta$ and which approximates $\{W(\theta)\}$ when $n$ is large. \textcolor{black}{In all our simulations, the nuisance parameters under the null model have been estimated via MLE and each simulated sample under $H_0$ is obtained via parametric bootstrap \citep{boostrap}.} We plot the results of our simulation in order to visualize the traces  of $\{W_n(\theta)\}$  as shown in Figure~\ref{upc_Gauss} for Example~1. (The analogous plots for Examples~2 and 3 appear in Figure~\ref{upc_others}.)
 In order to calculate \eqref{bound2}, it is important to provide an accurate estimate of $E[N_{c_0}]$. Hence,  we choose   $c_0$ to be at a level (on the y-axis) around which the process $\{W_n(\theta)\}$  oscillates   often, and thus, with respect to which the  upcrossings occur with high  frequency. 
 For Examples~1, 2 and 3, this leads to values $c_0$ equal to $0.1, 0.3$ and $0$, respectively.} Inspecting the smoothness of the trace plots also allows us to qualitatively assess Condition \ref{cond31} and verify the goodness of the approximation of $E[N_{c_0}]$ by $E[\tilde{N}_{c_0}]$, necessary for the validity of the results of Section  \ref{sec3}. 

As discussed  in Section~\ref{intro}, the implementation of our procedure requires the specification of a grid $\Theta_R$ over  $\Theta \equiv[\mathcal L; \mathcal U]$,  where $R$ is the number of times $H_0$ is tested versus the ensemble of sub-alternatives $H_{11},\dots,H_{1R}$. In practice, $R$ must either be chosen arbitrarily by the researcher or determined by the nature of the  experiment.
In either case,   $R$ must be sufficiently large to guarantee robustness of the results,  yet small enough to ensure computational efficiency when calculating \eqref{real_bound}. 
One possibility  is to choose $R$ large enough so that, for a given  $c_0$, $E[\tilde{N}_{c_0}]$ converges to a finite limit, which we expect, for sufficiently dense $\Theta_R$, to correspond to $E[N_{c_0}]$. This strategy requires us to set $c_0$ before setting $R$.

\textcolor{black}{In order to identify the value of $R$ that best negotiates the  trade-off between accuracy and computational efficiency,  one can consider different values 
of $R$ and for each of them compute an estimate of $E[{N}_{c_0}]$ by means of a small Monte Carlo simulation. The results can then be summarized in an \emph{upcrossing plot} where the  values for $R$ considered are reported on the $x$-axis and  the respective $\widehat{E[\tilde{N}_{c_0}]}$ estimates of $E[{N}_{c_0}]$ are reported on the $y$-axis. The upcrossing plot in
 the right panel of Figure~\ref{upc_Gauss} displays  Monte Carlo estimates $\widehat{E[\tilde{N}_{c_0}]}$ for the LRT in Example~1, under $H_0$, as a function of $R$ (with $R=15,30,50,100,200,500,1000$). For each value of $R$ considered, the grid points have been chosen to be equally spaced over $\Theta$. Analogous plots for Examples~2 and 3 appear in Figure \ref{upc_others}. For each $R$ considered  we computed 100 Monte Carlo simulations, each of size 1000. %The sample size of each simulation must be reasonably large to guarantee the asymptotic distribution of the sub-test statistics.
In all our examples, 100 simulations  are sufficient to achieve small Monte Carlo errors.}

\textcolor{black}{As a rule of thumb, if the number of upcrossings increases with $R$ but  does not converge, it means that the resolution is not sufficiently high to catch all the crossings or, the underlying process is not sufficiently smooth to guarantee $E[{N}_{c_0}]<\infty$. Conversely, if the number of upcrossings converges, 
as in the well-known scree-plot used for  Principal Component Analysis (PCA) \citep[e.g.,][p. 383]{statLearning}, we look for an ``elbow'' in the plot of $\widehat{E[\tilde{N}_{c_0}]}$.  The value of $R$ corresponding to the elbow is the smallest value for which  $\widehat{E[\tilde{N}_{c_0}]}$ converges to
 its limit, $E[{N}_{c_0}]$, up to  Monte Carlo error.  In physics terms, this  corresponds to the minimal value of $R$ for which $\widehat{E[\tilde{N}_{c_0}]}$ well approximates the number of upcrossings of the underlying continuous time process.
}

\begin{figure}[!h]
\begin{adjustwidth}{0cm}{0cm}
\begin{tabular*}{\textwidth}{@{\extracolsep{\fill}}@{}c@{}c@{}c@{}}
      \includegraphics[width=45mm]{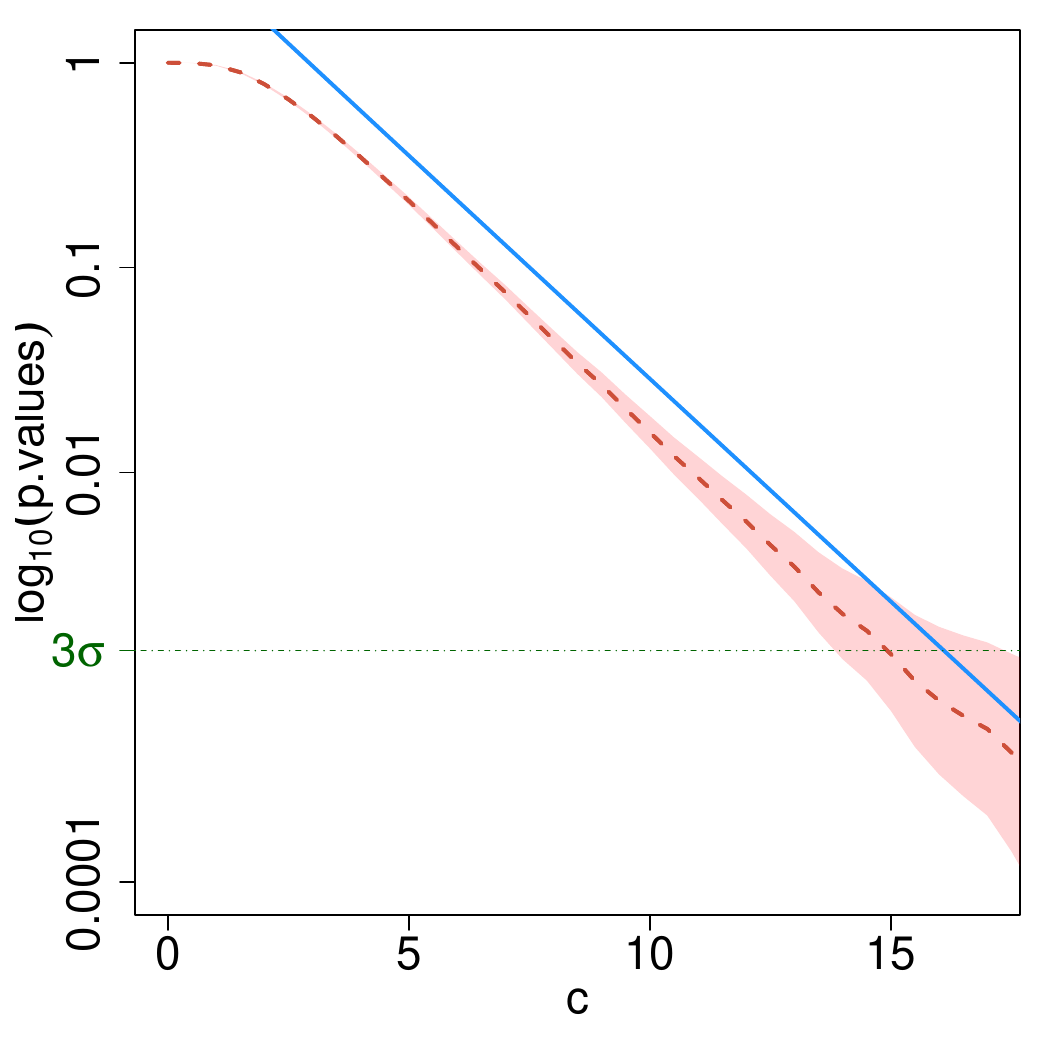} & \includegraphics[width=45mm]{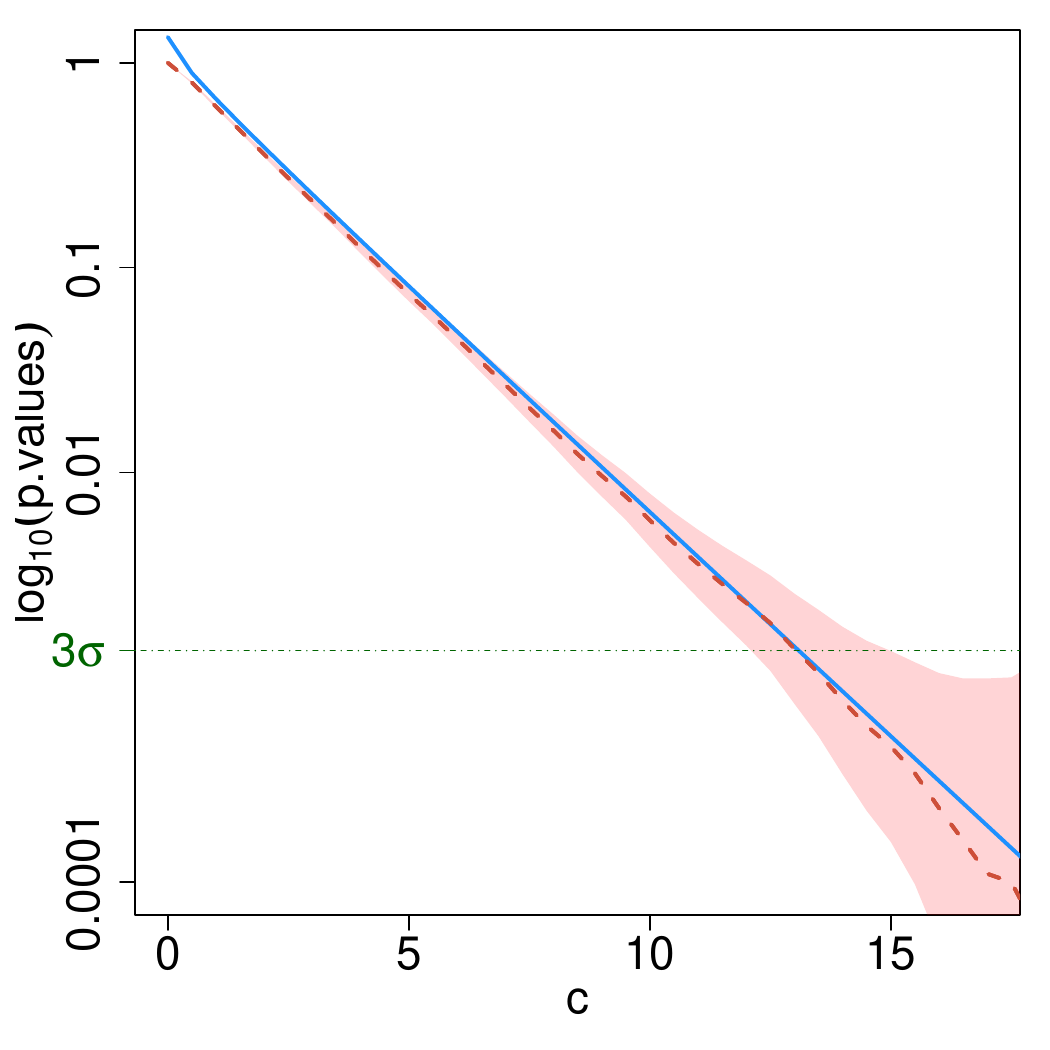}& \includegraphics[width=45mm]{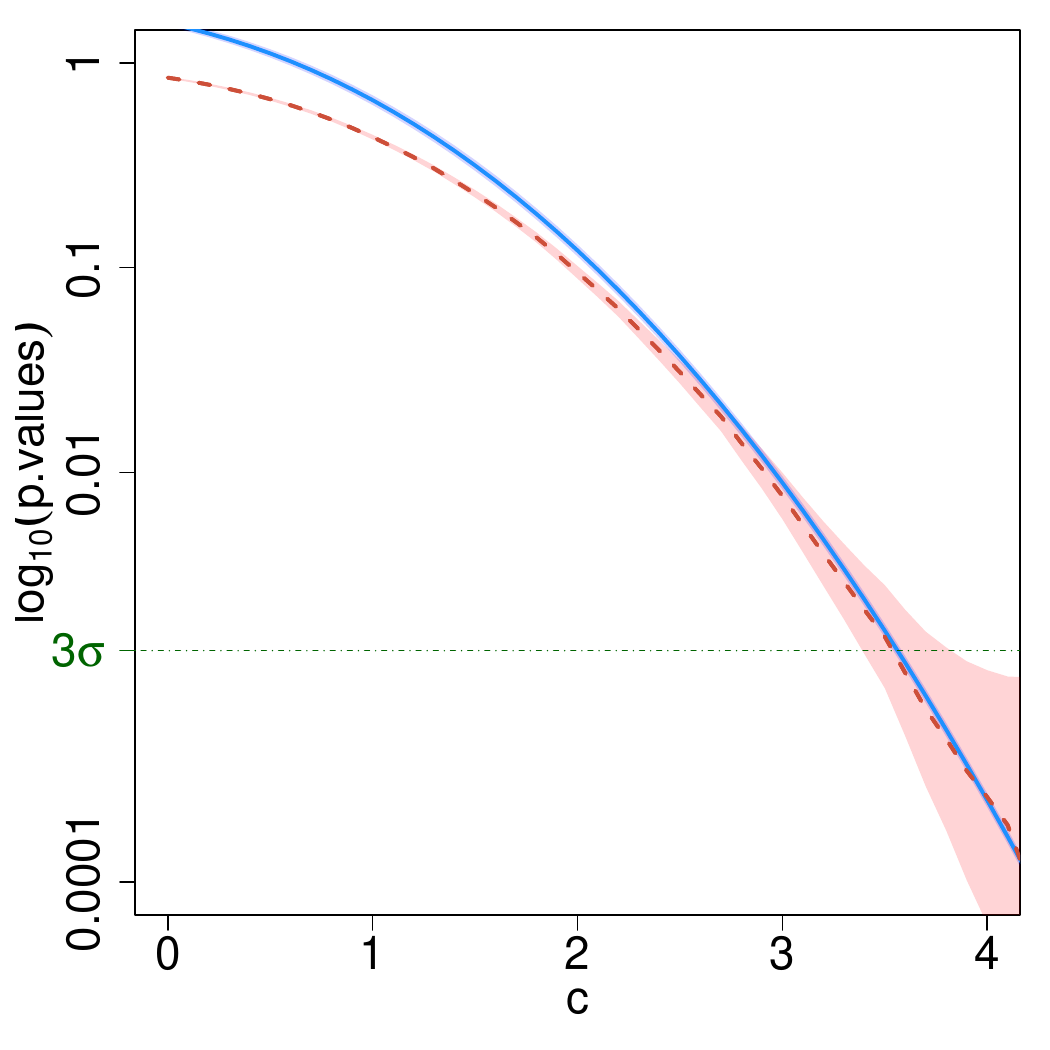} \\
\end{tabular*}
\end{adjustwidth}
\caption{Estimated   bound/approximation in \eqref{real_bound} (blue solid line), simulated global p-values (on $\log_{10}$-scale), Monte Carlo estimates of $P(\sup_{\theta \in \Theta}\{W(\theta)\}>c)$ (red dashed line),  and   Monte Carlo Errors (pink areas) for increasing values of the threshold $c$, for Example~1 (left panel), Example~2 (central panel) and Example~3 (right panel). Monte Carlo errors associated with 
$\widehat{E[\tilde{N}_{c_0}]}$ on the bound in \eqref{real_bound} are plotted in grey, but are too small to be visible.}
\label{assess}
\end{figure}
We also investigate the relationship between the width  of the signal in  the bump-hunting example,  and the grid resolution. In particular, we replicate the simulation for three choices of the Gaussian width, namely $\sigma =0.1\theta, \sigma =0.5\theta$ and $\sigma =\theta$. (In our actual analysis $\sigma=0.1\theta$.)  
As expected, wider signals correspond to smoother underlying processes (Figure~\ref{upc_Gauss}, left panel) and  $\widehat{E[\tilde{N}_{c_0}]}$ converges (Figure~\ref{upc_Gauss}, right panel)
at  lower grid resolution.

In general,  $R$  impacts the upper bound/approximation for the global p-value in \eqref{bound2}, as well as the observed value of the test statistics, $c_R$, which we assume  converges to  $c$,  as $R\rightarrow\infty$. Specifically, if the gap between $\theta_r$ and $\theta_{r+1}$ is wider than the signal width, $c_R$ may underestimate $c$,  and the signal may be missed. Thus, if  the signal is suspected to be localized over a small region of the search interval, a higher resolution is required to accurately estimate \eqref{real_bound} and avoid false negatives, which would in turn adversely affect the power of the test.

\textcolor{black}{
Conversely, in Examples~2 and 3, the signal  is spread either over the whole parameter space or over a large portion of it. 
In these cases the choice of $R$ should  be based on the desired level of accuracy of  both $c_R$ as an estimate for the maximum of the underlying process and of the  value of $\theta$ at which the maximum occurs, i.e.,
\begin{equation}
\label{thetatilde}
\tilde{\theta}= \text{argmax}_{\theta_r \in \Theta_R} \{W(\theta_r)\}.
\end{equation}}

Finally, based on the elbow in the upcrossing plots in Figures \ref{upc_Gauss} and \ref{upc_others}, the values of $R$ we select are $R=100$ in Example~1 (with $\sigma=0.1\theta$ as in \eqref{ex1}), $R=50$ in Example~2, and $R=30$ in Example~3.  In order to guarantee accuracy of at least $0.5$ for the identified location, $\tilde{\theta}$, of the break-point, however, we  set $R=50$  in Example~3. For each of the models considered, we computed   \eqref{real_bound} using the  $R$ and $c_0$ selected above.
 The results obtained are compared in Figure~\ref{assess} with the Monte Carlo  estimates of  $P(\sup_{\theta \in \Theta}\{W(\theta)\}>c)$ for increasing values of $c$, obtained  using 100,000 simulations, each of size 10,000.
The pink areas correspond to the respective Monte Carlo errors.
The Monte Carlo errors associated to   the estimate $\widehat{E[\tilde{N}_{c_0}]}$  for $E[\tilde{N}_{c_0}]$ in \eqref{real_bound} (and displayed on a lower scale in the upcrossing plots) are also incorporated in Figure~\ref{assess}, but they are too small to be visible. As expected, the estimated TOHM bounds approach the ``truth'' as $c\rightarrow \infty$. Convergence appears to be slower for Example~1. The plots, however, are presented on $\log_{10}$-scale, and thus in all cases we obtain a good approximations of the global p-values.

\section{Comparing TOHM and Bonferroni's bounds}
\label{analysis}
In fields such as high energy physics and astrophysics, experiments are often  characterized by the search of one signal over a wide pool of possibilities. The simplest possible way to tackle this problem using classical  Multiple Hypothesis Testing (MHT)  is by means of Bonferroni correction \citep{bonferroni35, bonferroni36}. The Bonferroni bound for the global p-value is  
\begin{equation}
\label{bonfcorr1}
p_{\mathrm BF}=R\cdot \min_{\theta_r \in \Theta_R} P(W(\mathcal L)\geq w(\theta_r))=R\cdot P(W(\mathcal L)\geq c_R).
\end{equation}
\textcolor{black}{The standard Bonferoni correction, $p_{\mathrm BF}$, used to bound statistical significance in multiple testing also yields a bound on  $P(\max_{\theta_r \in \Theta_R}\{W(\theta_r)\}\geq c_R)$. Specifically,
\begin{align*}
P\biggl(\max_{\theta_r \in \Theta_R} \{W(\theta_r)\}\geq c_R\biggl)&=P\biggl(\cup_{\theta_r \in \Theta_R}\{W(\theta_r)>c_R\}\biggl)\leq \sum_{\theta_r \in \Theta_R}P(W(\theta_r)>c_R)\\
&=R\cdot P(W(\mathcal{L})>c_R)=p_{BF}.
\end{align*}}

\begin{figure}[!h]
\begin{adjustwidth}{0cm}{0cm}
\begin{center}
\hspace{-1.5cm}
\begin{tabular*}{\textwidth}{@{\extracolsep{\fill}}@{}c@{}c@{}c@{}}
 \includegraphics[width=55mm,height=57mm]{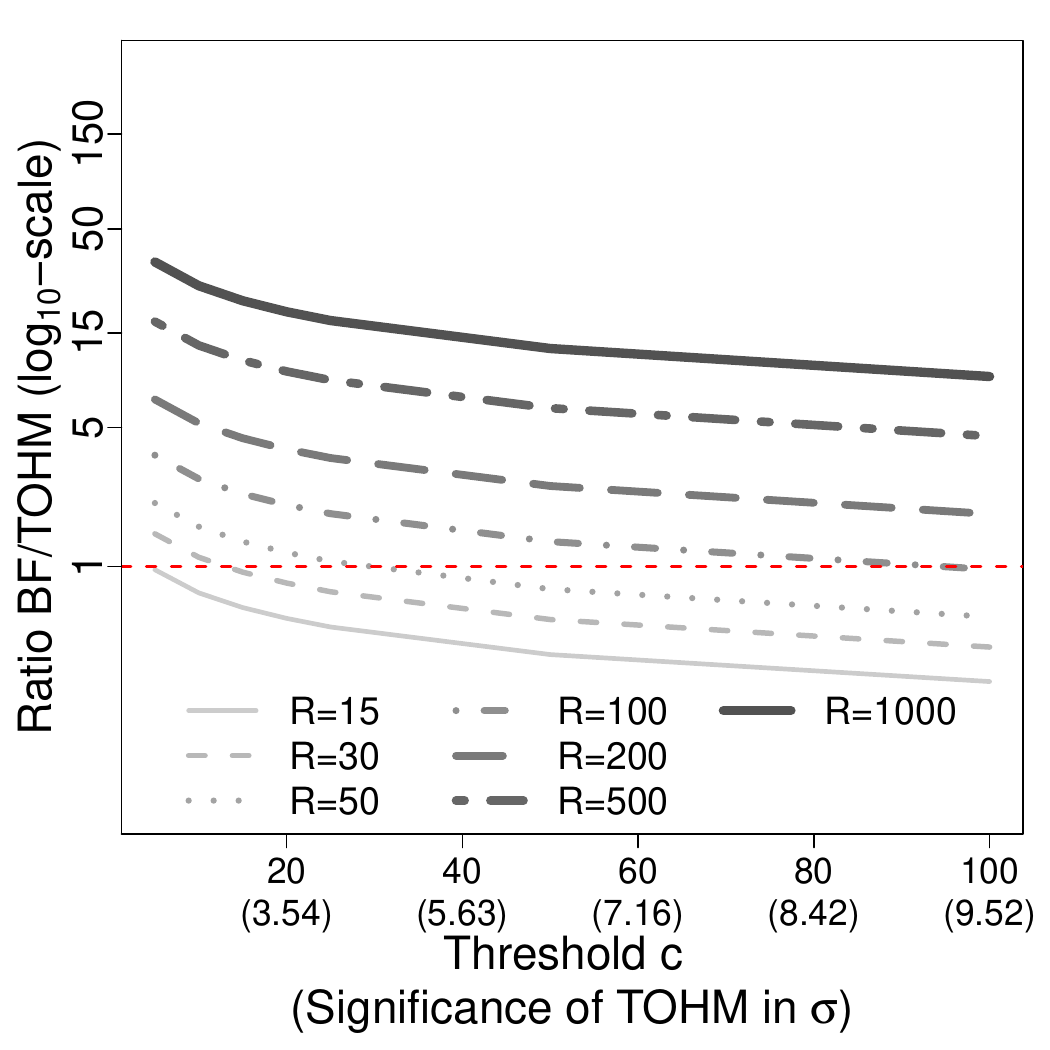} & \hspace{-0.25cm}  \includegraphics[width=55mm,height=57mm]{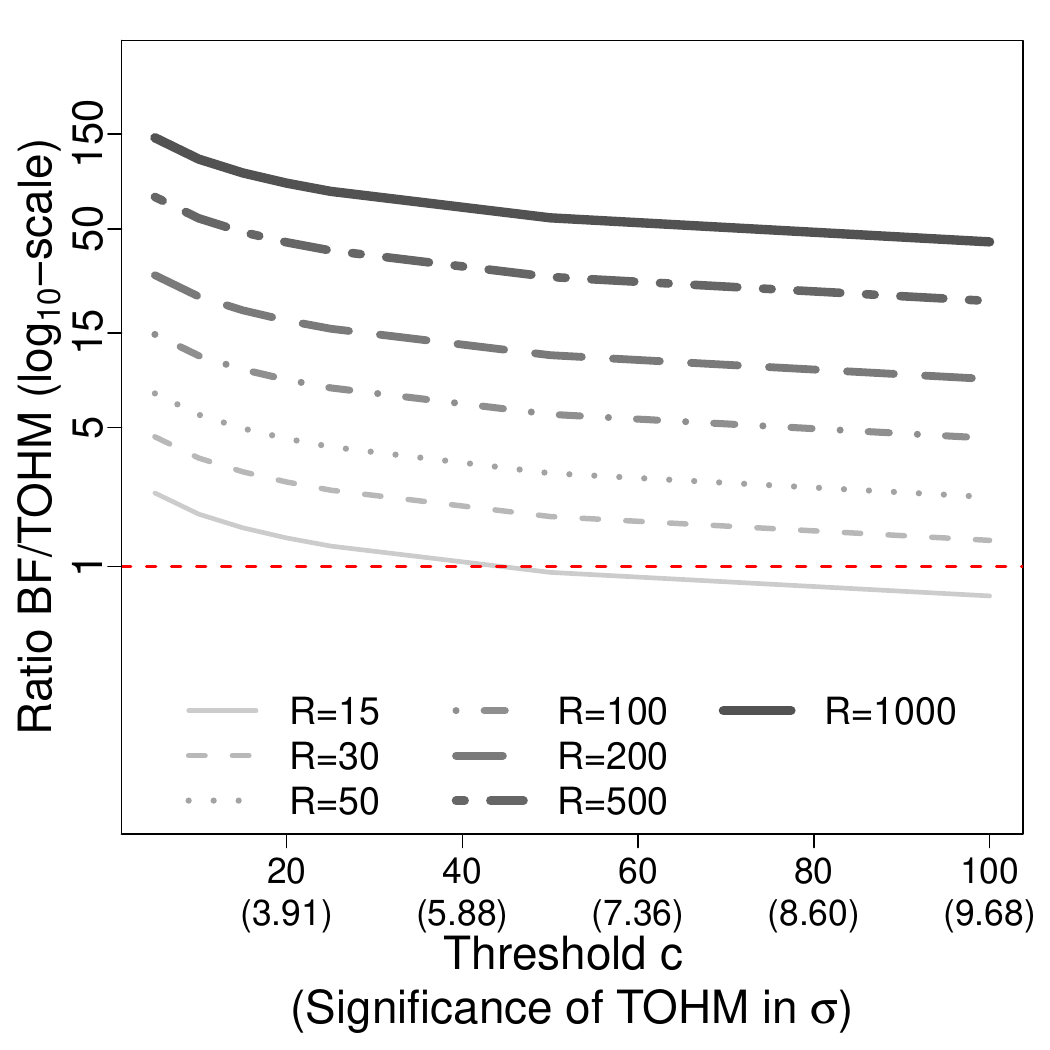}  &  \hspace{-0.25cm} \includegraphics[width=55mm,height=57mm]{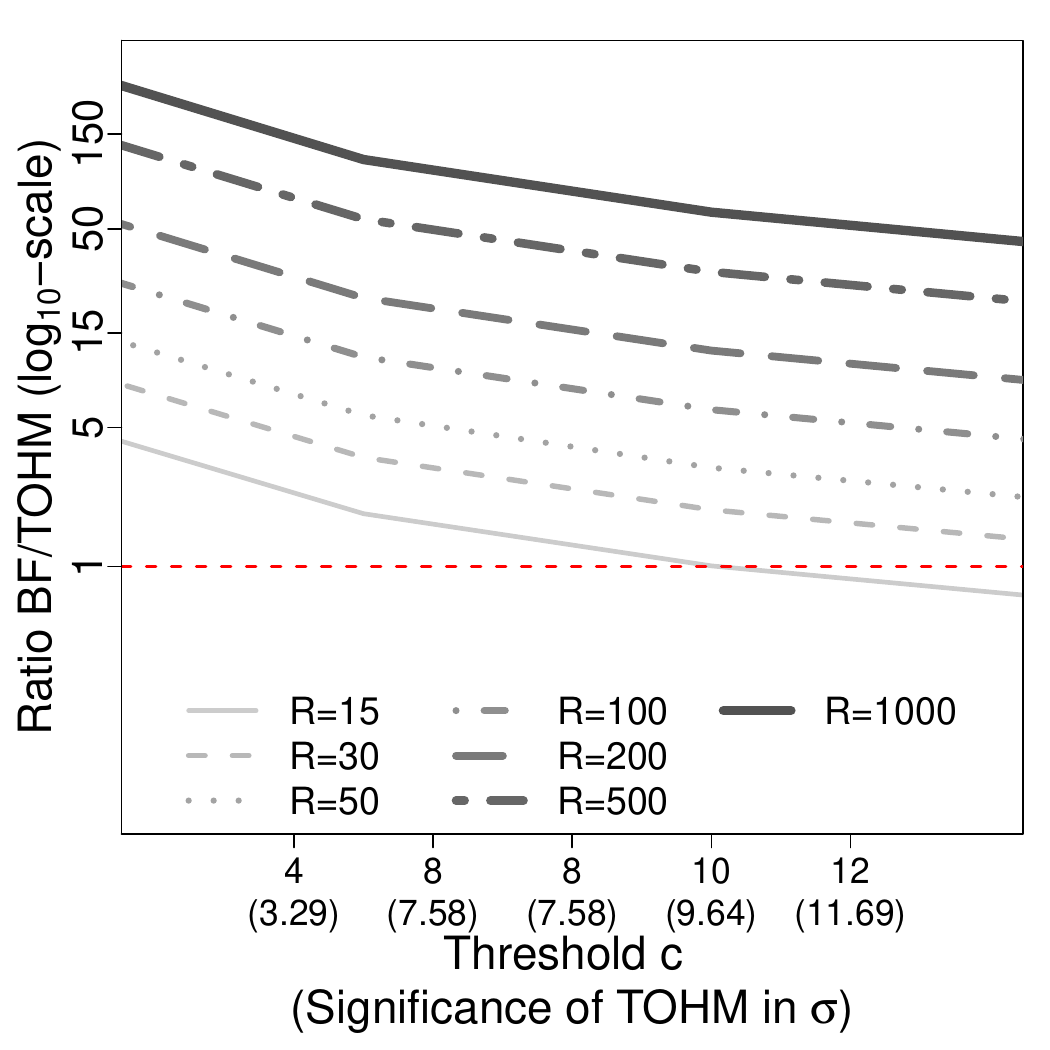}\\
\end{tabular*}
\end{center}
\end{adjustwidth}
\caption[Comparing TOHM and Bonferroni's bounds]{Ratio of Bonferroni and TOHM's bounds   at increasing values of  $c$ (and corresponding  significance for TOHM), and considering different resolutions (grey curves).  The left, central and right panels correspond to Example 1,  2 and 3, respectively.  }
\label{EVTBonf}
\end{figure} 

In this section, we investigate  the relationship between the TOHM and Bonferroni bounds using simple constructs from EVT in order to individuate situations where the latter can be used without leading to overly conservative results.

First, we introduce the distinction between  upcrossings  and  \emph{exceedances} of $\{W(\theta_r)\}$.  Specifically, an exceedance of $c_R$ by $\{W(\theta_r)\}$ occurs at $\theta_r$ if   $\{W(\theta_r)>c_R\}$.  
 An  illustration of the difference between upcrossings and exceedances is given in Figure~\ref{exceedances}. 
\textcolor{black}{We denote   by  $\tilde{N}_{c_R}$, the  process of  exceedances  of $c_R$ by $\{W(\theta_r)\}$, and let $\dot{N}_{c_R}$ be the process of upcrossings as defined in \ref{Ncrdef}.} Notice that 
\begin{align}
\label{derivationBF1}
 E[\dot{N}_{c_R}]&= \sum^{R}_{r=1}P\biggl(W(\theta_r)\geq c_R\biggl)=  \sum^{R}_{r=1}P\biggl(W(\theta_r)\geq \max_{\theta_{r'} \in \Theta_R}\{w(\theta_{r'})\}\biggl)\\
\label{derivationBF2}
&= R \min_{\theta_r \in \Theta_R} P(W(\mathcal L)\geq w(\theta_r))=p_{\mathrm BF}
\end{align}

\textcolor{black}{
Because each upcrossing requires at least one exceedance, $E[\dot{N}_{c_R}]\geq E[\tilde{N}_{c_R}]$. Moreover,  we  expect that the clusters of exceedances corresponding to each upcrossing to   be  smaller, and consequently  $E[\dot{N}_{c_R}]$ to approach $E[\tilde{N}_{c_R}]$ as $c_R$ increases. 
$E[\dot{N}_{c_R}]$ can be easily computed using $p_{\mathrm BF}$ in \eqref{derivationBF1}-\eqref{derivationBF2}; whereas, when   $\{W(\theta)\}$ satisfies  Condition \ref{cond31}, $E[\tilde{N}_{c_R}]$ is approximately equal to the second term in  \eqref{real_bound}, for large $R$.  Further,  $E[\tilde{N}_{c_R}]$ dominates the first term in \eqref{real_bound}, as $c_R\rightarrow \infty$. Thus, it is natural to consider if there are  situations where $\eqref{real_bound}$ and $p_{\mathrm BF}$ are approximately equivalent bounds on $P(\max_{\theta_r \in \Theta_R}\{W(\theta_r)\}\geq c_R)$, i.e,
\begin{equation}
\label{approximation}
P(W(\mathcal L)>c_R)+ \frac{a(c_R)}{a(c_0)}E[\tilde{N}_{c_0}] \approx  p_{\mathrm BF},
\end{equation}
for $c_0\leq c_R$, $c_R\rightarrow +\infty$ and $R\rightarrow +\infty$.
Unfortunately, simultaneously quantifying the  rates at which $c_R$ and $R$ must increase for \eqref{approximation} to hold is not an easy task; hence, we investigate the approximation in \eqref{approximation} by means of a numerical simulation where we compare the  performance of Bonferroni and the TOHM  bounds with respect to the number of tests considered and the level of significance
for Examples 1, 2 and 3. }

\textcolor{black}{
 The results are reported in Figure \ref{EVTBonf}, where we plot the ratio of the two bounds for increasing values of $c$, using different grid sizes, $R$.  Because the signed-root LRT, $\{Q_n(\theta)\}$, is used in Example 3 rather than the LRT, smaller values of $c$ correspond to equally significant results. In  the horizontal axes, the statistical significance   is reported in terms of  $\sigma$-significance, i.e., the number of standard deviations from the mean of a standard normal distribution that corresponds to the tail probability expressed by the one-sided p-value, i.e.,
\[\#\sigma=\Phi^{-1}(1-\text{p-value}),\]
where $\Phi$ is the standard normal cumulative function.}

\textcolor{black}{
In Examples 2 and  3,  Bonferroni is  always more conservative than the TOHM bound when at least 30 tests are performed. For $R=15$, Bonferroni becomes less conservative only when the level of significance achieved is of the order of $6\sigma$ and $11\sigma$, respectively.}

\textcolor{black}{A more interesting situation is observed for Example 1. Here, equivalence of $p_{\mathrm TOHM}$ and $p_{\mathrm BF}$   occurs for values of $c$ much smaller than those for which the same limit is achieved in Examples 2 and 3.
Further, when  $R\leq50$, Bonferroni quickly becomes  less conservative than the TOHM bound as $c$ increases. For $R=50$ for instance, Bonferroni performs better than TOHM when $c>30$ ($\sim4.5\sigma$ significance).}

\textcolor{black}{
Finally,   all the plots in   Figure \ref{EVTBonf} suggest that the TOHM bound  is  preferable to Bonferroni with very high resolutions, i.e. $R\geq500$, for all the significance levels considered (up to $\sim 10\sigma$).}

\textcolor{black}{It is important to point out that  the  value of $R$  selected  via the upcrossing plots discussed in Section \ref{choosingR} is the minimum number of grid points (among those considered) for which  $\widehat{E[\tilde{N}_{c_0}]}$ converges to its limit. As $R$ increases beyond this point, 
 the estimated TOHM bound remains constant, whereas Bonferroni's continues to increase. This implies that, when the number of tests to be conducted can be selected arbitrarly,  Bonferroni will not be overly conservative if the ``elbow'' in the upcrossings plot appears at a relatively small  value of $R$ and the observed value of $c$  is large. However, practitioners should keep in mind that when attempting to identify the signal location, $\tilde{\theta}$, a higher resolution is typically required and thus TOHM is preferable.}

\begin{table}[!h]
\fontsize{9}{11}\selectfont{
 \centering
\begin{tabular}{c|c|ccccc}
\noalign{\global\arrayrulewidth0.08cm}
 \hline
 \noalign{\global\arrayrulewidth0.08pt}
 &  &   &  &  &  &  \\
{Example} &{Test}& {Method}  &{ $R$ }&{$c_R$}&                 $\tilde{\theta}$              &{p-value }  \\
                                           &&                                              &                                  &                                        &   &                  (Significance)                                 \\
 &  &   &  &  &  &  \\
\noalign{\global\arrayrulewidth0.08cm}
 \hline
 \noalign{\global\arrayrulewidth0.08pt}
 &  &   &  &  &  &  \\
\multirow{2}{*}{Example~1}&$H_0:\eta=0$&Bonferroni & \multirow{2}{*}{100}& \multirow{2}{*}{38.326}  & \multirow{2}{*}{3.404 }   &$2.99\cdot 10^{-8}$ ($5.42\sigma$)\\ 
 & $H_1:\eta>0$&  TOHM  &  &  &  & $2.11\cdot 10^{-8}$ ($5.48\sigma$)\\
 &  &   &  &  &  &  \\
\hline
 &  &   &  &  &  &  \\
\multirow{4}{*}{Example~2}&$H_0:\eta=0$&Bonferroni & \multirow{2}{*}{50}& \multirow{2}{*}{21.021}  & \multirow{2}{*}{27.265 }   &$1.14\cdot 10^{-4}$ ($3.69\sigma$)\\ 
 & $H_1:\eta>0$& TOHM  &  &  &  & $2.51\cdot 10^{-5}$ ($4.06\sigma$)\\
&$H_0:\eta=1$&Bonferroni & \multirow{2}{*}{50}& \multirow{2}{*}{0.606 }  & \multirow{2}{*}{ 27.890 }   &$>1$ ($0.00\sigma$ )\\ 
 & $H_1:\eta<1$&  TOHM  &  &  &  & $ 7.201\cdot 10^{- 1}$ ($0.58 \sigma$)\\
 &  &   &  &  &  &  \\
\hline
 &  &   &  &  &  &  \\
\multirow{2}{*}{Example~3}&$H_0:\xi=0$&Bonferroni & \multirow{2}{*}{50}& \multirow{2}{*}{11.826}  & \multirow{2}{*}{31.266 }   &$1.43\cdot 10^{-30}$ ($11.43\sigma$)\\ 
 & $H_1:\xi\neq0$& TOHM  &  &  &  & $5.06\cdot 10^{-31}$ ($11.52\sigma$)\\
 &  &   &  &  &  &  \\
\noalign{\global\arrayrulewidth0.05cm}
 \hline
 \noalign{\global\arrayrulewidth0.05pt}
\end{tabular}
\vspace{0.1cm}
\caption{Summary of the results of TOHM and MHT via Bonferroni on real data for Examples~1,  2 and 3.}
\label{real_table}}
\end{table}

\subsection{Data analyses}
\label{application}
In this section we compare the TOHM and  Bonferroni bounds  for Examples 1, 2 and 3. The results are summarized in Table \ref{real_table}. 
In the dark matter search problem of Example 1, we obtain  a significance in favour of the presence of a dark matter emission of about $5.4\sigma$ using both TOHM and MHT.  
This result is not surprising since $c_R=38.326$ and as shown
 in  the central panel of Figure \ref{EVTBonf},  at $c\approx40$ the gray line associated with $R=100$ is very close the red dashed line.
The signal location selected  is close to the truth (3.5GeV), and  the estimated model is plotted as a solid red line in the left panel of Figure \ref{real_plots}; the  signal location selected, $\tilde{\theta}=3.404$, is indicated by the green dotted vertical line.

In Example 2 both TOHM and  Bonferroni reject the hypothesis that the observed emission is due to a power-law distributed cosmic source at $4.06\sigma$ and $3.69\sigma$ respectively.
Because this example involves a non-nested models comparison,  we invert the null of the hypotheses in order to avoid meaningless results  (see Section~\ref{examples} for more details). In the inverted test, the power-law model cannot be rejected. Both the fitted dark matter model and the fitted power-law cosmic source model are displayed in the central panel of Figure \ref{real_plots}. In Example 2, when testing \eqref{testex1}, the value of $\theta$ (i.e., the signal annihilation of the dark matter model)  selected by  TOHM   is $\tilde{\theta}=27.265$GeV. This is somewhat off from the true value used to simulate the data ($\theta=35$GeV), perhaps because 
our analysis does not account for  instrumental errors. Our analysis also only uses the spectral  energy of the $\gamma$-ray signals, whereas in practice the directions of the $\gamma$-ray would also be used, thus increasing the statistical power.

\textcolor{black}{
Finally, for the break-point regression model in Example 3, both TOHM and MHT give similar inferences  ($11.52\sigma$ and $11.43\sigma$ respectively) when rejecting the hypothesis of a linear model with no break-point. \textcolor{black}{The equivalence among the two procedure is likely  due to the very high statistical significance, and the only moderately large number of tests conducted ($R=50$).} 
The fitted model is displayed in Figure \ref{real_plots} where the green triangle corresponds to the optimal break-point location, i.e., the maximum of the signed-root LRT process occurs at a  mother's age of 31.266 years.}

%%%%%%%%%%%%%%%%Section 6

\section{Discussion}
\label{discussion}
In this paper we discuss a  highly generalizable method to efficiently conduct  statistical tests  under non-standard conditions, including bump-hunting, structural change detection and non-nested models comparison.  

The main advantages of the method proposed are its easy implementation and its efficiency in providing accurate inference, while controlling for very small Type~I errors rates.
Following \citet{davies87} and \citet{gv10} we combine the theoretical framework of EVT with the practical simplicity of Monte Carlo simulations and  we  generalize their results beyond the LRT and $\chi^2$. Using a suite 
of simulation studies we show that as few as  100 Monte Carlo simulations are often sufficient to achieve a high level of accuracy. 
 Although we do not investigate  the power of TOHM here, readers interested in power are directed to \citet{davies77} for a formal derivation of lower and upper bounds of the power function in the normal case, or the simulation studies conducted in \citet{algeri16} and \citet{algeri2} for the $\bar{\chi}^2_{01}$ case. 

\textcolor{black}{From a more practical perspective, we  propose   simple graphical tools to select the threshold $c_0$ and  to specify an appropriate number of sub-tests $R$ to guarantee robustness of the resulting inference. Finally, we investigate the relationship between the TOHM and Bonferroni bounds and we implement both procedures on our running examples.} 
Extensions of our results to the case where the nuisance parameter  specified only under the alternative, ${\bm \theta}$, is multi-dimensional are the subject of a forthcoming paper \citep{algeri18}. 

It is important to point out that the stringent significance requirements play a critical role in both the theory discussed in Section \ref{sec3} and practical applications. Specifically, this setup is particularly well suited for searches in high energy physics, where the significance level necessary to claim a discovery is of at least $5\sigma$. However, in light of the recent ``p-value crisis'', culminated with the  \emph{Journal Basic and Applied Social Psychology} banning the use of the p-value in future submissions \citep{ASAstatement, nature}, stringent significance criteria may become more popular in other scientific communities.

\vskip 14pt
\noindent {\large\bf Supplementary Materials}
In Section S.1 we discuss the error rate of  \eqref{bound2} for Gaussian, $\chi^2$ and $\bar{\chi}^2_{01}$ processes. Proofs of Result 2  and  Result 3 are collected in Section S.2. Additionally figures are reported in  Section S.3. Data used in Examples~1 and 2 are also downloadable among the Supplementary Materials.
\par
%%%%%%%%%%%%%%%%%%%%%%%%%%%%%%%%%%%%%%%%%%%%%%%%%%%%%%%%%%%%%%%%%%%%%%%%%%%%%%%%%%%%%%%%%%%%%%%%%%%%%%%%%%%%%%%%%%%%%%%%%%%%
\vskip 14pt
\noindent {\large\bf Acknowledgements}
\textcolor{black}{The authors thank two anonymous referees and the associate editor for their constructive feedback}. SA and DvD also thank Jan Conrad for the valuable discussion of the physics problems which motivated this work, and  Brandon Anderson who provided the Fermi-LAT datasets used in the analyses.
 DvD acknowledges support from Marie-Skodowska-Curie RISE (H2020-MSCA-RISE-2015-691164) Grant  provided by the European Commission. 
\par

\bibliographystyle{chicago}      % Chicago style, author-year citations
\bibliography{biblioBio2}  

\begin{thebibliography}{}

\bibitem[\protect\citeauthoryear{Adler}{Adler}{2000}]{adler2000}
Adler, R. (2000).
\newblock On excursion sets, tube formulas and maxima of random fields.
\newblock {\em Annals of Applied Probability\/}, 1--74.

\bibitem[\protect\citeauthoryear{Adler and Taylor}{Adler and
  Taylor}{2009}]{adlerbook}
Adler, R. and J.~Taylor (2009).
\newblock {\em Random fields and geometry}.
\newblock Springer Science \& Business Media.

\bibitem[\protect\citeauthoryear{Algeri et~al.}{Algeri et~al.}{2016}]{algeri2}
Algeri, S. et~al. (2016).
\newblock On methods for correcting for the look-elsewhere effect in searches
  for new physics.
\newblock {\em Journal of Instrumentation\/}~{\em 11\/}(12), P12010.

\bibitem[\protect\citeauthoryear{Algeri, Conrad, and van Dyk}{Algeri
  et~al.}{2016}]{algeri16}
Algeri, S., J.~Conrad, and D.~van Dyk (2016).
\newblock A method for comparing non-nested models with application to
  astrophysical searches for new physics.
\newblock {\em Monthly Notices of the Royal Astronomical Society:
  Letters\/}~{\em 458\/}(1), L84--L88.

\bibitem[\protect\citeauthoryear{Algeri and van Dyk}{Algeri and van
  Dyk}{2018}]{algeri18}
Algeri, S. and D.~van Dyk (2018).
\newblock Testing one hypothesis multiple times: The multidimensional case.
\newblock {\em arXiv:1803.03858\/}.

\bibitem[\protect\citeauthoryear{Anderson et~al.}{Anderson
  et~al.}{2016}]{refB3}
Anderson, B. et~al. (2016).
\newblock Search for gamma-ray lines towards galaxy clusters with the
  fermi-lat.
\newblock {\em Journal of Cosmology and Astroparticle Physics\/}~{\em
  2016\/}(02), 026.

\bibitem[\protect\citeauthoryear{Andrews}{Andrews}{1993}]{andrews93}
Andrews, D. (1993).
\newblock Tests for parameter instability and structural change with unknown
  change point.
\newblock {\em Econometrica: Journal of the Econometric Society\/}, 821--856.

\bibitem[\protect\citeauthoryear{Andrews and Ploberger}{Andrews and
  Ploberger}{1994}]{andrews94}
Andrews, D. and W.~Ploberger (1994).
\newblock Optimal tests when a nuisance parameter is present only under the
  alternative.
\newblock {\em Econometrica: Journal of the Econometric Society\/}, 1383--1414.

\bibitem[\protect\citeauthoryear{Atkinson}{Atkinson}{1970}]{atkinson}
Atkinson, A. (1970).
\newblock A method for discriminating between models.
\newblock {\em Journal of the Royal Statistical Society. Series B
  (Methodological)\/}, 323--353.

\bibitem[\protect\citeauthoryear{Atwood et~al.}{Atwood et~al.}{2009}]{atwood}
Atwood, W.~B. et~al. (2009).
\newblock The large area telescope on the fermi gamma-ray space telescope
  mission.
\newblock {\em The Astrophysical Journal\/}~{\em 697\/}(2), 1071.

\bibitem[\protect\citeauthoryear{Bergstr{\"o}m, Ullio, and
  Buckley}{Bergstr{\"o}m et~al.}{1998}]{bergstrom}
Bergstr{\"o}m, L., P.~Ullio, and J.~Buckley (1998).
\newblock Observability of $\gamma$ rays from dark matter neutralino
  annihilations in the milky way halo.
\newblock {\em Astroparticle Physics\/}~{\em 9\/}(2), 137--162.

\bibitem[\protect\citeauthoryear{Bonferroni}{Bonferroni}{1935}]{bonferroni35}
Bonferroni, C. (1935).
\newblock {Il calcolo delle assicurazioni su gruppi di teste}.
\newblock In {\em Studi in Onore del Professore Salvatore Ortu Carboni}, pp.\
  13--60. Rome.

\bibitem[\protect\citeauthoryear{Bonferroni}{Bonferroni}{1936}]{bonferroni36}
Bonferroni, C. (1936).
\newblock Teoria statistica delle classi e calcolo delle probabilit\`{a}.
\newblock {\em Pubblicazioni del R Istituto Superiore di Scienze Economiche e
  Commerciali di Firenze\/}~{\em 8}, 3--62.

\bibitem[\protect\citeauthoryear{Chernoff}{Chernoff}{1954}]{chernoff}
Chernoff, H. (1954).
\newblock On the distribution of the likelihood ratio.
\newblock {\em The Annals of Mathematical Statistics\/}, 573--578.

\bibitem[\protect\citeauthoryear{Choudalakis}{Choudalakis}{2011}]{bumpphy}
Choudalakis, G. (2011).
\newblock {On hypothesis testing, trials factor, hypertests and the
  BumpHunter}.
\newblock In {\em {Proceedings, PHYSTAT 2011. ArXiv:1101.0390}}.

\bibitem[\protect\citeauthoryear{Cox}{Cox}{1962}]{cox62}
Cox, D. (1962).
\newblock Further results on tests of separate families of hypotheses.
\newblock {\em Journal of the Royal Statistical Society. Series B
  (Methodological)\/}, 406--424.

\bibitem[\protect\citeauthoryear{Cox}{Cox}{2013}]{cox13}
Cox, D. (2013).
\newblock A return to an old paper:‘tests of separate families of
  hypotheses’.
\newblock {\em Journal of the Royal Statistical Society: Series B (Statistical
  Methodology)\/}~{\em 75\/}(2), 207--215.

\bibitem[\protect\citeauthoryear{Cram{\'e}r and Leadbetter}{Cram{\'e}r and
  Leadbetter}{2013}]{cramer}
Cram{\'e}r, H. and M.~Leadbetter (2013).
\newblock {\em Stationary and related stochastic processes: Sample function
  properties and their applications}.
\newblock Courier Corporation.

\bibitem[\protect\citeauthoryear{Davies}{Davies}{1977}]{davies77}
Davies, R. (1977).
\newblock Hypothesis testing when a nuisance parameter is present only under
  the alternative.
\newblock {\em Biometrika\/}~{\em 64\/}(2), 247--254.

\bibitem[\protect\citeauthoryear{Davies}{Davies}{1987}]{davies87}
Davies, R. (1987).
\newblock Hypothesis testing when a nuisance parameter is present only under
  the alternative.
\newblock {\em Biometrika\/}~{\em 74\/}(1), 33--43.

\bibitem[\protect\citeauthoryear{Davies}{Davies}{2002}]{davies02}
Davies, R. (2002).
\newblock Hypothesis testing when a nuisance parameter is present only under
  the alternative: linear model case.
\newblock {\em Biometrika\/}, 484--489.

\bibitem[\protect\citeauthoryear{Efron and Tibshirani}{Efron and
  Tibshirani}{1994}]{boostrap}
Efron, B. and R.~Tibshirani (1994).
\newblock {\em An introduction to the bootstrap}.
\newblock CRC press.

\bibitem[\protect\citeauthoryear{Gross and Vitells}{Gross and
  Vitells}{2010}]{gv10}
Gross, E. and O.~Vitells (2010).
\newblock Trial factors for the look elsewhere effect in high energy physics.
\newblock {\em The European Physical Journal C\/}~{\em 70\/}(1-2), 525--530.

\bibitem[\protect\citeauthoryear{Hansen}{Hansen}{1991}]{hansen91}
Hansen, B. (1991).
\newblock Inference when a nuisance parameter is not identified under the null
  hypothesis.
\newblock {\em Rochester Center for Economic Research Working Paper No. 296\/}.

\bibitem[\protect\citeauthoryear{Hansen}{Hansen}{1992a}]{hansen92}
Hansen, B. (1992a).
\newblock The likelihood ratio test under nonstandard conditions: testing the
  markov switching model of gnp.
\newblock {\em Journal of applied Econometrics\/}~{\em 7\/}(S1).

\bibitem[\protect\citeauthoryear{Hansen}{Hansen}{1992b}]{hansen92b}
Hansen, B. (1992b).
\newblock Testing for parameter instability in linear models.
\newblock {\em Journal of policy Modeling\/}~{\em 14\/}(4), 517--533.

\bibitem[\protect\citeauthoryear{Hansen}{Hansen}{1996}]{hansen96}
Hansen, B. (1996).
\newblock Inference when a nuisance parameter is not identified under the null
  hypothesis.
\newblock {\em Econometrica: Journal of the econometric society\/}, 413--430.

\bibitem[\protect\citeauthoryear{Hansen}{Hansen}{1999}]{hansen99}
Hansen, B. (1999).
\newblock Threshold effects in non-dynamic panels: Estimation, testing, and
  inference.
\newblock {\em Journal of econometrics\/}~{\em 93\/}(2), 345--368.

\bibitem[\protect\citeauthoryear{Hotelling}{Hotelling}{1939}]{hotelling39}
Hotelling, H. (1939).
\newblock Tubes and spheres in n-spaces, and a class of statistical problems.
\newblock {\em American Journal of Mathematics\/}~{\em 61\/}(2), 440--460.

\bibitem[\protect\citeauthoryear{James et~al.}{James
  et~al.}{2013}]{statLearning}
James, G. et~al. (2013).
\newblock {\em An introduction to statistical learning}, Volume 112.
\newblock Springer.

\bibitem[\protect\citeauthoryear{Leek and Peng}{Leek and Peng}{2015}]{nature}
Leek, J. and R.~Peng (2015).
\newblock Statistics: P values are just the tip of the iceberg.
\newblock {\em Nature\/}~{\em 520\/}(7549), 612.

\bibitem[\protect\citeauthoryear{Lyons}{Lyons}{2013}]{lyons2015}
Lyons, L. (2013).
\newblock Discovering the significance of 5 sigma.
\newblock {\em arXiv preprint arXiv:1310.1284\/}.

\bibitem[\protect\citeauthoryear{Muggeo et~al.}{Muggeo
  et~al.}{2008}]{segmented}
Muggeo, V. et~al. (2008).
\newblock Segmented: an r package to fit regression models with broken-line
  relationships.
\newblock {\em R news\/}~{\em 8\/}(1), 20--25.

\bibitem[\protect\citeauthoryear{Quandt}{Quandt}{1974}]{quandt}
Quandt, R. (1974).
\newblock A comparison of methods for testing nonnested hypotheses.
\newblock {\em The Review of Economics and Statistics\/}, 92--99.

\bibitem[\protect\citeauthoryear{Rice}{Rice}{1944}]{rice}
Rice, S. (1944).
\newblock Mathematical analysis of random noise.
\newblock {\em Bell Labs Technical Journal\/}~{\em 23\/}(3), 282--332.

\bibitem[\protect\citeauthoryear{Self and Liang}{Self and Liang}{1987}]{self}
Self, S. and K.-Y. Liang (1987).
\newblock Asymptotic properties of maximum likelihood estimators and likelihood
  ratio tests under nonstandard conditions.
\newblock {\em Journal of the American Statistical Association\/}~{\em
  82\/}(398), 605--610.

\bibitem[\protect\citeauthoryear{Taylor and Adler}{Taylor and
  Adler}{2003}]{taylor2003}
Taylor, J. and R.~Adler (2003).
\newblock Euler characteristics for gaussian fields on manifolds.
\newblock {\em Annals of Probability\/}, 533--563.

\bibitem[\protect\citeauthoryear{Taylor and Worsley}{Taylor and
  Worsley}{2008}]{taylor2008}
Taylor, J. and K.~Worsley (2008).
\newblock Random fields of multivariate test statistics, with applications to
  shape analysis.
\newblock {\em The Annals of Statistics\/}, 1--27.

\bibitem[\protect\citeauthoryear{van Dyk}{van Dyk}{2014}]{dvd}
van Dyk, D. (2014).
\newblock The role of statistics in the discovery of a higgs boson.
\newblock {\em Annual Review of Statistics and Its Application\/}~{\em 1},
  41--59.

\bibitem[\protect\citeauthoryear{Wasserstein and Lazar}{Wasserstein and
  Lazar}{2016}]{ASAstatement}
Wasserstein, R. and N.~Lazar (2016).
\newblock The asa's statement on p-values: context, process, and purpose.

\end{thebibliography}

\vskip .65cm
\noindent
 \textcolor{black}{
Sara Algeri\\
School of Statistics\\
University of Minnesota \\
Minneapolis, MN, 55455 
\vskip 2pt
\noindent
E-mail: salgeri@umn.edu
\vskip 2pt
\vspace{2cm}
\noindent
David van Dyk\\
Statistics Section\\
Dept of Mathematics\\
Imperial College London\\
London, UK SW7 2AZ
\vskip 2pt
\noindent
E-mail: d.van-dyk@imperial.ac.uk
}

\end{document}